\documentclass[journal]{IEEEtran}
\usepackage[table,xcdraw]{xcolor}
\usepackage[utf8]{inputenc}
\usepackage{amsmath}
\usepackage{graphicx}
\usepackage{refstyle}
\usepackage[table,xcdraw]{xcolor}
\usepackage{subcaption}
\usepackage{tikz}
\usepackage{booktabs}
\usepackage{multirow}
\usepackage{cite}

\begin{document}
\title{Detection and localization of Coronary Arterial Lesion with the Aid of Impedance Cardiography}

\author{Sudipta Ghosh,~\IEEEmembership{} Bhabani Prasad Chattopadhyay, Ram Mohan Roy,  Jayanta Mukherjee,~\IEEEmembership{} and Manjunatha Mahadevappa,~\IEEEmembership{}
	
\thanks{S. Ghosh is with School of Medical Science and Technology, Indian Institute of Technology Kharagpur, Kharagpur, West Bengal-721302, INDIA, e-mail: sudipta.ghosh@iitkgp.ac.in, sudipta.kanti@gmail.com}
\thanks{B.P. Chattopadhyay and R.M. Roy is with Department of Cardiology, Medical College and Hospital, Kolkata, West Bengal-700073, INDIA, e-mail: bhabani07@yahoo.co.in \& dr.roy99@gmail.com}
\thanks{J. Mukherjee is with Department of Computer Science and Engineering, Indian Institute of Technology Kharagpur, Kharagpur, West Bengal-721302, INDIA, e-mail: jay@cse.iitkgp.ac.in}
\thanks{M. Mahadevappa is with School of Medical Science and Technology, Indian Institute of Technology Kharagpur, Kharagpur, West Bengal-721302, INDIA, e-mail: mmaha2@smst.iitkgp.ac.in}}


\maketitle

\begin{abstract}
In recent years, coronary artery disease is escalating and is likely to assume an epidemic proportion by 2030. Currently the reliable methods for detection of coronary arterial lesions are either conventional coronary angiogram (CAG) or MDCT (Multiple Detector Computed Tomography) coronary angiogram. Conventional CAG is an invasive procedure. Conventional CAG and CT (Computed Tomography) angiogram, both require expert supervision of either an interventional cardiologist or a radiologist. In this work, we have proposed a novel design and method for non-invasive detection and localization of coronary arterial lesion using Impedance Cardiography (ICG). The ICG signal recorded by the proposed device is used to extract feature points and compute augmentation index, amplitude and other time related parameters. The extracted features are used as input to a trained artificial  neural  network,  for  detection and prediction of coronary arterial lesions. The trained network generates specialized models, to be used for diagnosis of arterial lesions. The proposed methodology detects lesion in Left main coronary artery (LMCA), Left anterior descending artery (LAD), Diagonal branch, Left circumflex artery (LCX), and Right coronary artery (RCA) with an accuracy of 92\%, 82\%, 76\%, 76\%, 84\% respectively. The proposed device could be also used by a common individual for detection of arterial lesion without any expert supervision, unassisted. The proposed algorithm eliminates the need of CAG for diagnosis of coronary arterial lesions (stenosis), and provides an insight into a new method for non-invasive monitoring of cardiovascular haemodynamics, detection and localization of coronary arterial lesion.
\end{abstract}

\begin{IEEEkeywords}
Impedance Cardiography, Augmentation Index, Artery lesion, Coronary Angiogram, Artificial Neural Network
\end{IEEEkeywords}

\IEEEpeerreviewmaketitle

\section{Introduction}
\IEEEPARstart{C}{ardiovascular} diseases (CVD) have seen an increasing trend in recent years. World Health Organisation (WHO) projects cardiovascular diseases as the primary cause of death, globally. According to WHO there have been 17.7 million deaths due to CVDs in 2015 alone, which pertains to 31\% of total global deaths \cite{who2017}. Another report by the American Heart association, suggests that by 2030 the number of deaths due to CVDs will rise to 23.6 million \cite{benjamin2017heart}. This ever increasing occurence of CVDs across the world, combined with the frugal nature of cardiovascular care in ICU (Intensive Care Unit) patients, calls for development of better techniques for detection and assessment of cardiovascular diseases. Cardiovascular diseases are basically disorders of heart and its associated blood vessels. CVDs include but are not limited to, coronary heart disease, peripheral arterial disease, rheumatic heart disease, congenital heart disease, deep vein thrombosis, cerebrovascular disease, and pulmonary embolism \cite{who2017}. Hemodynamic monitoring which encompasses the study of development and propagation of flow and pressure pulses, including but not limited to: systemic, pulmonary arterial and venous pressures and cardiac output, have traditionally been utilized by the clinicians. However, they are costly and require on-site expert supervision \cite{bigatello2002hemodynamic}.

The first and foremost thing that comes to one's mind regarding the health of one's heart, and its associated vessels is the stiffness of the vessels. Essentially arteries are supposed to be pulsatile in nature, but with age and other factors they tend to become stiff and loose their pulsatile nature \cite{liu2015arteries}. The other factors may include plaque, cholesterol deposition, congenital defects, etc. The stiffer the arteries are, the worse is the condition of the patient, which leads to cardiovascular diseases. The stiffness of the arteries could be considered as a precursor of CVDs. Studies have been conducted to find the relevance of arterial stiffnes as a cardiological risk factor. Researchers have employed pulse wave velocity (PWV) to study arterial stiffness \cite{townsend2015recommendations}, \cite{gao2014improved}, \cite{zhang2011radial}. Studies have been conducted using two pulse synthesis model of PPG waveform, to determine the risk of coronary artery diseases due to altered arterial stiffness \cite{goswami2011relevance}.  Calculation of PWV is highly dependent on the accurate and precise measurement of pulse transit time (PTT). Accurate and precise measurement of PTT involves two types of blood pressure waveform, one at the central and another at the peripheral arterial site. PTT calculation also involves calculation of foot to foot delay of these two acquired blood pressure waveforms \cite{vardoulis2013validation} \cite{papaioannou2014validation}. This sort of PWV and PTT measurement requires an expert supervision, in order to simultaneously record the two waveforms. The recording usually is done with the help of hand held tonometers. PTT is used for ambulatory pressure monitoring \cite{ding2016continuous}, \cite{mukkamala2015toward}. It has also been used to estimate transfer functions for calculation of central blood pressure waveforms \cite{westerhof2008individualization} \cite{swamy2009adaptive}, \cite{hahn2012subject}, \cite{gao2016simple}. Researchers have also proposed a two pulse synthesis model for non-invasive derivation of important arterial parameters, that could lead to monitoring of human vascular health \cite{goswami2010new,deb2007cuff}. Commercially available devices like, SphygmoCor\textsuperscript{TM} (AtCor Medical, Sydney, Australlia), use electrocardiograph (ECG) in order to determine PWV \cite{butlin2016large}. PTT has also been used to measure systolic blood pressure, and to estimate changes in blood pressure during obstetric spinal anaesthesia \cite{ahlstrom2005noninvasive}, \cite{sharwood2005assessment}. 

Methods based on PTT and PWV measure the stiffness index of arteries, but are not able to localise the occurence of lesion. Presently clinicians perform Coronary angiogram (CAG), and radiologists perform MDCT angiogram for localization and detection of coronary arterial lesion. Conventional CAG is an invasive procedure. It requires  supervision by an interventional cardiologist,  and involves complications of an interventional surgical procedure \cite{stiver2017complete}. MDCT coronary angiogram needs expertise of a radiologist trained in cardiac CT. As mentioned above most of the hemodynamic parameter monitoring techniques are invasive in nature. In 1966 Kubicek \textit{et al.} \cite{kubicek1966development} proposed a novel non-invasive technique, Impedance Cardiography (ICG) for measuring cardiac output and body fluid composition. Impedance cardiography (ICG) measures the ionic conduction of human body in contrast to electrical conduction characteristics of human \cite{nyboer1950electrical}, \cite{yamakoshi1980noninvasive}. The volume of arterio-venous blood within a specific segment of human body is deemed responsible for the static and transient values of electrical conductivity. The variation in impedance ($\Delta Z$) obtained due to the pulsatile, peripheral blood flow of limbs has been mathematically related to the pulsatile change in volume by Nyboer \cite{nyboer1950electrical}, as shown in Eq. \ref{eq1}.

\begin{equation}
		\centering
		{\Delta V = \left( {\frac{{\rho {L^2}}}{{{Z_0}^2}}} \right)\Delta Z}
		\label{eq1}
\end{equation}

where $\Delta V$ is the pulsatile volume change with respect to resistivity, $\rho$ = resistivity of blood in $\Omega$-cm, L= length of the explored limb in cm, $Z_{0}$= measured impedance at minimum volume, and $\Delta Z$ is the amount of variation in impedance.

In impedance cardiography an alternating current is injected into a particular segment of the human body, where it conducts mostly through blood in comparison to the more resistive parts, like bone, fat etc. As the injected current flows through blood, there is a change in electric potential due to the pulsatile changes in blood flow. This change in electrical potential is measured and its corresponding impedance change is observed. If $R_{b}$ is the total resistance, $R_{0}$ is the static resistance and $R_{n}$  is the alternate resistance of a particular segment of human body, the change incurred in resistance $\Delta R$, due to the pulsatile blood flow is given by Eq. \ref{eq2}.
\begin{equation}
		\centering
		{R_b} = \frac{{R{}_0{R_n}}}{{{R_0} - {R_n}}} = \frac{{{R_0}{R_n}}}{{\Delta R}}
		\label{eq2}
\end{equation}

In this work, we propose a novel technique for non-invasive detection and localization of arterial lesion by processing ICG and using an artificial neural network based classifier. The results obtained from the proposed technique were compared with the observations drawn by the physicians using coronary angiogram on the same set of subjects. The latter data set is considered as the `gold standard' data for verification and validation of the proposed technique. In our previous work we proposed a basic and simplifed circuit for obtaining ICG signal from the fore-arm of a human body \cite{ghosh2016electrical}. In this study we have also presented an improvised version of the device reported earlier.

\section{Materials and Methodology}
\subsection{Subject Population}
The present study is being performed at the department of cardiology, ``Medical College and Hospital, Kolkata". A total of 50 subjects have been recruited in this study so far. Each of the recruited subjects' electrocardiograph (ECG) and echocardiogram results are also recorded. Requisite ethical clearance for the study has been obtained from the institute ethical committe of ``Indian Institute of Technology Kharagpur" and  ``Medical College and Hospital, Kolkata". The baseline characteristics of the recruited subjects are shown in \ref{table1}. Subjects with confirmed cases of cardiovascular disease, undergoing coronary angiogram, and having abnormal ECG were recruited in this current study, whereas the exclusion criteria are listed below:
\begin{itemize}
	\item The subject should not have any electronic device, like pacemaker, implanted.
	\item The subject should not be below 18 years of age.
	\item The subject should not be suffering from septic shock.
	\item The subject should not have aortic valve regurgitation and defect on septum.
	\item The subject should not suffer from tachycardia, with heart rate greater than 200 beats per minute.
	\item The subject should not be underweight (i.e. below 30 kg) and also should not be more than 155 kg in weight.
\end{itemize}
\begin{center}
	\begin{table}[!t]
		\centering
		\caption{Baseline characteristics of the study group}
		\label{table1}
		\resizebox{\columnwidth}{!}{%
			\begin{tabular}{|
					>{\columncolor[HTML]{FFFFFF}}l |
					>{\columncolor[HTML]{FFFFFF}}l |
					>{\columncolor[HTML]{FFFFFF}}l |}
				\hline
				\textbf{Number of Subjects (n)} & \textbf{50}  \\ \hline
				\textbf{Gender (Male/ Female)} & 43 / 7  \\ \hline
				\textbf{Age (years)} & 51.37 $\pm$ 7.93  \\ \hline
				\textbf{Height (cm)} & 164.84 $\pm$ 7.89  \\ \hline
				\textbf{Weight (kg)} & 64.18 $\pm$ 0.07  \\ \hline
				\textbf{BMI} & 23.65 $\pm$ 2.72  \\ \hline
				\textbf{Body Surface Area (\textbf{$m^2$)}} & 1.71 $\pm$ 0.12 \\ \hline
				\textbf{Heart Rate (beats per minute)} & 89.64 $\pm$ 6.88  \\ \hline
				\textbf{Systolic Pressure (mm of Hg)} & 126.84 $\pm$ 19.86 \\ \hline
				\textbf{Diastolic Pressure (mm of Hg)} & 81.90 $\pm$ 9.29  \\ \hline
				\textbf{Left Ventricular Ejection Fraction (\%)} & 53.32 $\pm$ 14.57  \\ \hline
			\end{tabular}%
		}
	\end{table}
\end{center}

\subsection{ICG circuit implementation and feature extraction}
\begin{figure*}[htbp]
	\centering{\includegraphics[width=\textwidth]{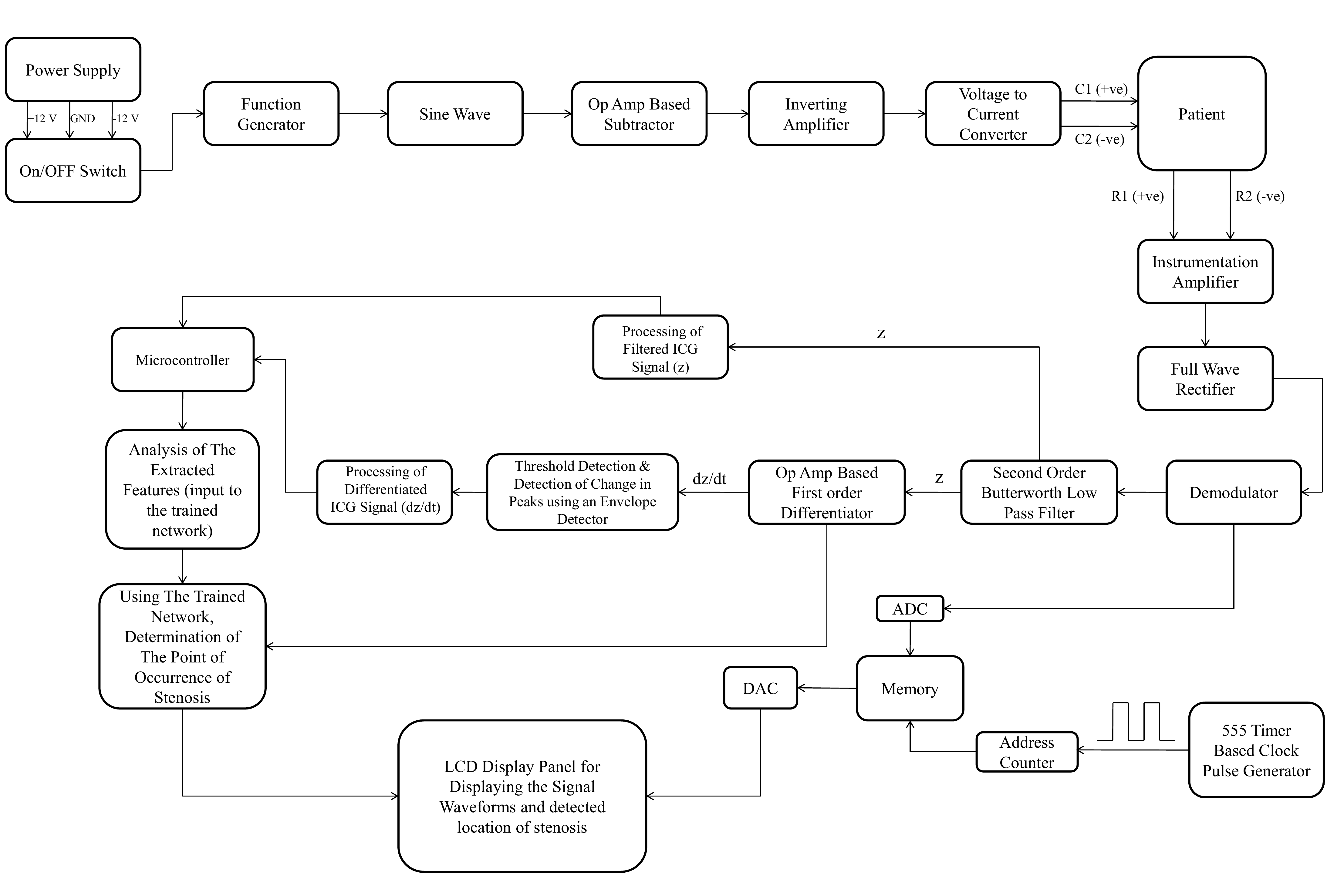}}	
	\caption{Block diagram of the implemented circuit}
	\label{fig1}
\end{figure*}

The authors have designed a cost effective, easy to use impedance cardiography (ICG) device for recording ICG signals from the subjects. The basic block diagram of the implemented circuit is shown in Fig. \ref{fig1}. The device is designed for use on the subject's forearm, with the aid of a tetrapolar configuration. The device uses a function generator to generate a biphasic sinusoidal wave of 50 \% duty cycle. The generated signal is found to be oscillating about a value much higher than D.C zero, so the generated sine wave is fed into a subtractor in order to receive a signal which oscillates about D.C zero. The generated signal has to have 50 \% duty cycle such that there is no charge accumulation on the skin surface of the subject. Charge accumulation has to be zero so that there is no risk of skin burn to the subject. The signal thus generated is then amplified and fed into a voltage to current converter. The voltage to current converter is designed in a floating ground configuration. The output of the voltage to current converter is then fed into two excitation electrodes, placed on the forearm of the subject. The specification of the excitation pulse injected through the excitation electrodes are as follows:
\begin{itemize}
	\item Frequency: 15 kHz
	\item Current: 3 mA
	\item Source type: Sine Wave
\end{itemize}

\begin{figure}[htbp]
	\centering{\includegraphics[width=\columnwidth]{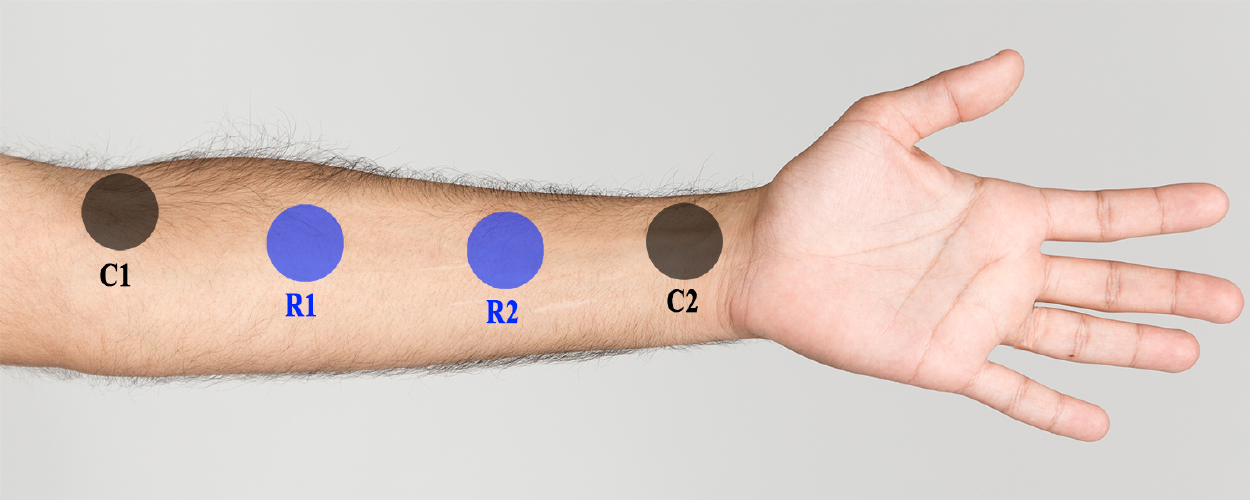}}	
	\caption{Electrode placement on the fore-arm of the subject}
	\label{fig2}
\end{figure}
\begin{figure}[htbp]
	\centering{\includegraphics[width=\columnwidth]{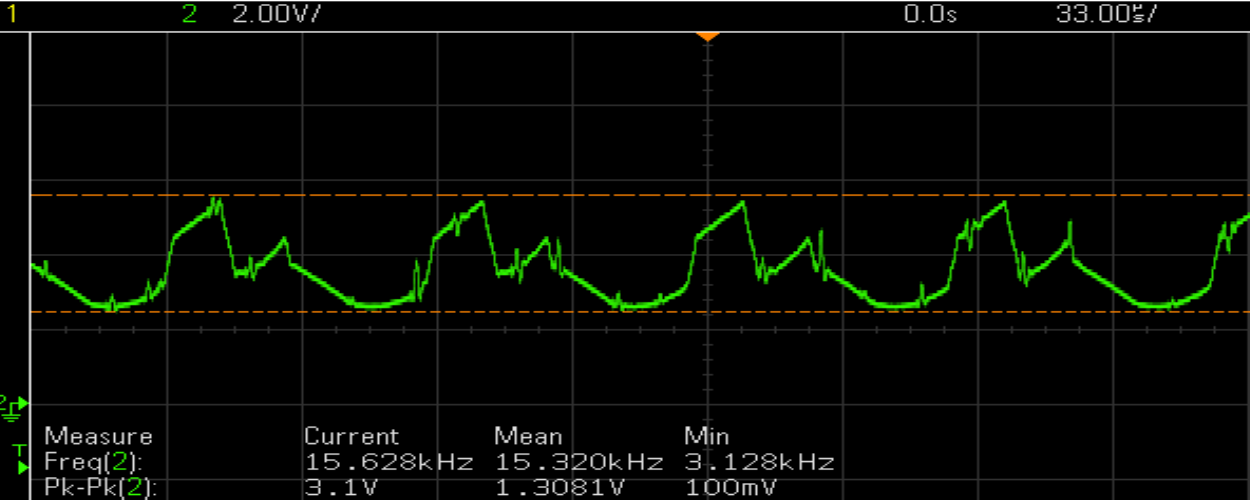}}	
	\caption{Actual ICG signal recorded before filtering (raw signal)}
	\label{fig3}
\end{figure}
\begin{figure}[htbp]
	\centering{\includegraphics[width=\columnwidth]{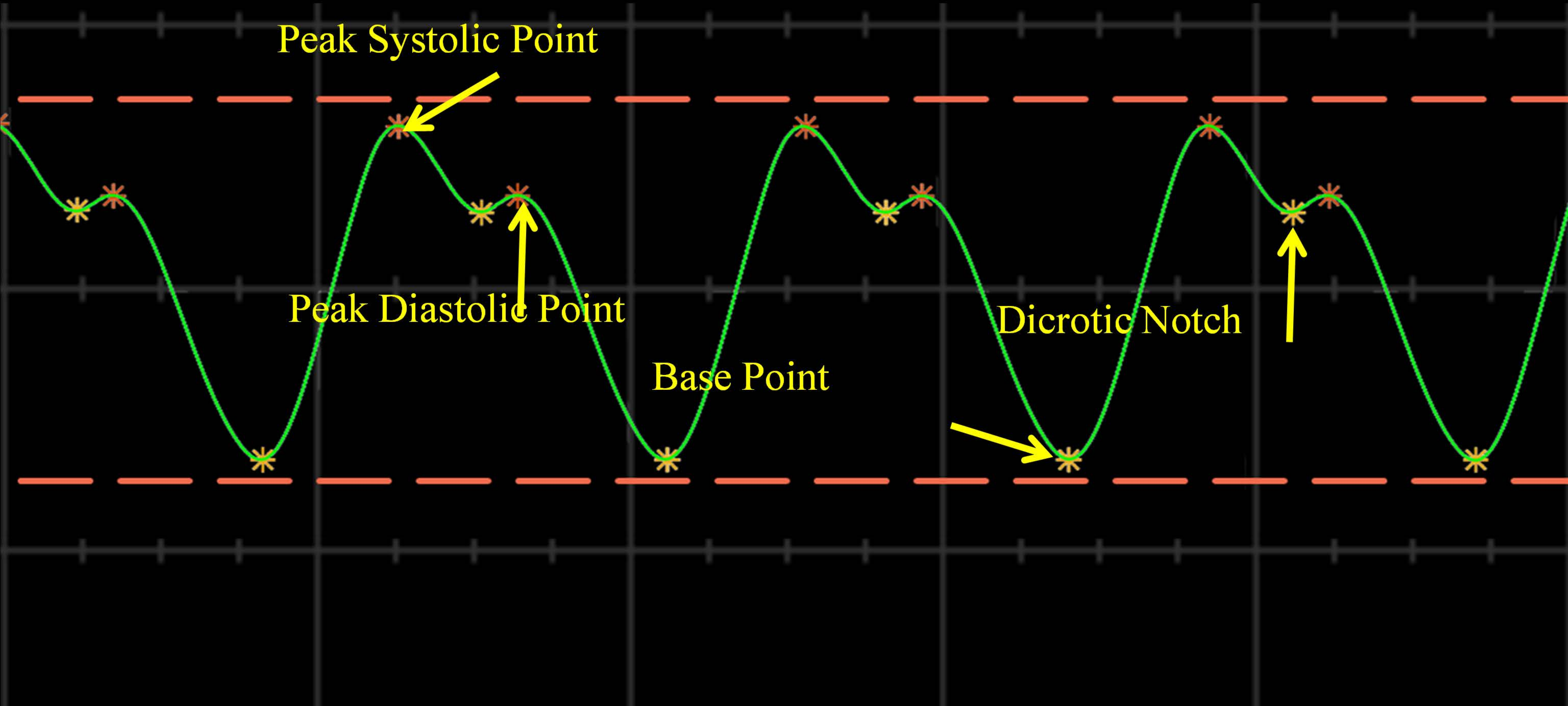}}	
	\caption{Filtered ICG signal obtained after passing the raw signal through a low pass butterworth filter}
	\label{fig4}
\end{figure}
\begin{figure*}[htbp]
	\centering{\includegraphics[width=0.75\textwidth]{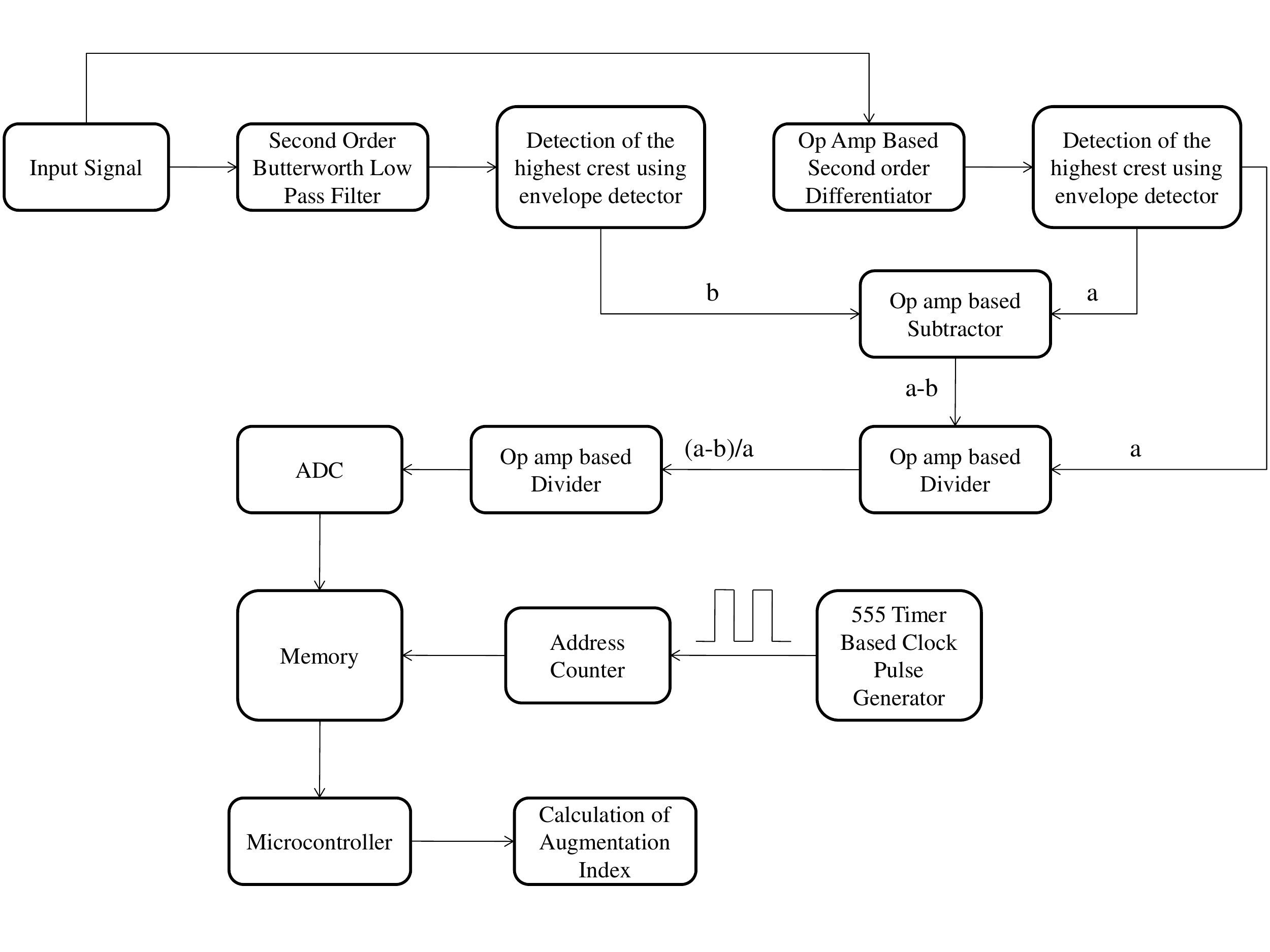}}	
	\caption{Block diagram of the circuit used for calculation of augmentation index (\textit{AIx}) from the differentiated signal (dz/dt)}
	\label{fig5}
\end{figure*}

The voltage sensing electrodes are placed in between the two excitation electrodes. The positioning of the electrodes is shown in Fig. \ref{fig2}, where R1, R2 are the voltage sensing electrodes, and C1, C2 are the excitation electrodes. The voltage signal received by the electrodes R1 and R2 are fed into an instrumentation amplifier, which is further rectified using a full wave rectifier. A demodulator is used to remove the 15 kHz carrier wave from the sensed signal. Once the signal is demodulated, it is passed through a second order low pass butterworth filter, with a cut off frequency of 250 Hz. The raw signal obtained by the sensing electrodes, and the filtered signal are shown in Fig. \ref{fig3} and Fig. \ref{fig4} respectively. 
The filtered signal is then fed into a differentiator circuit from where the differentiated signal is again processed. An overview of the steps employed in the processing of the differentiated signal are shown in Fig. \ref{fig5}.

\begin{figure*}[htbp]
	\centering{\includegraphics[width=0.75\textwidth]{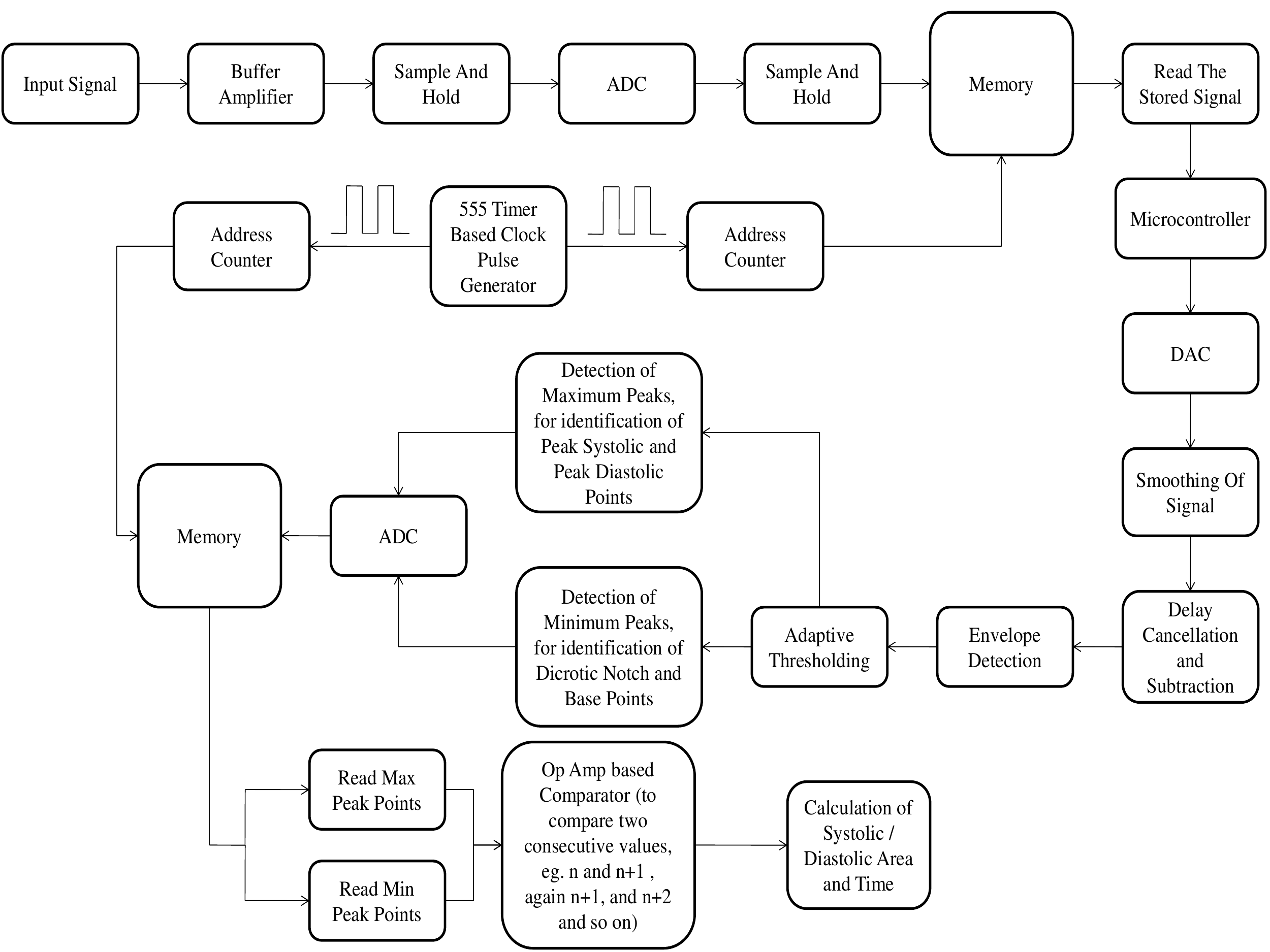}}	
	\caption{Block diagram of the circuit used for processing the filtered ICG signal z}
	\label{fig6}
\end{figure*}

The filtered signal is also further processed using envelope detector and an adaptive thresholding in order to identify the best available cardiac cycles. This signal is also used to identify peak systolic point, peak diastolic point, dicrotic notch and base point as shown in Fig. \ref{fig4}. The main steps and the circuitory employed for processing of the raw and filtered signal are shown in Fig. \ref{fig6}. 

Before moving on to the last stage, wherein the systolic/ diastolic area and time is calculated (as shown in Fig. \ref{fig6}), the maximum values and the minimum values has already been identified, and the same has been stored in a memory. The maximum values stored, correspond either to the peak systolic points or the peak diastolic points. The minimum values stored, correspond either to the dicrotic notch or the base points. At first the array containing the maximum values, is called. The comparator compares the values stored only in that array in an incremental fashion. It compares the n\textsuperscript{th} value with (n+1)\textsuperscript{th} value. If the n\textsuperscript{th} value is more than the (n+1)\textsuperscript{th}  value, then it is stored as the peak systolic point, otherwise it is stored as the peak diastolic point. Thereby when the last comparison is over, all the peak systolic values and the peak diastolic values of the entire recorded signal is saved. A similar process is followed for the array containing the minimum values. The comparator compares the values stored only in that array in an incremental fashion. It compares the n\textsuperscript{th} value with (n+1)\textsuperscript{th} value. If the n\textsuperscript{th} value is more than the (n+1)\textsuperscript{th}  value, then it is stored as the dicrotic notch, otherwise it is stored as the base point. Thereby when the last comparison is over, all the dicrotic notch values and the base point values of the entire recorded signal is saved.

\begin{figure}[htbp]
	\centering{\includegraphics[width=\columnwidth]{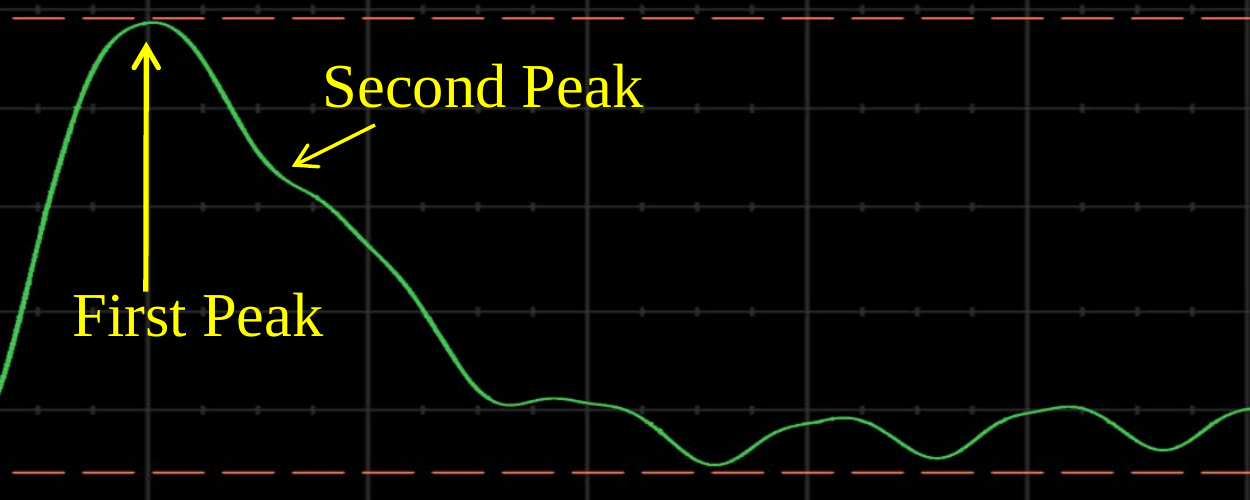}}	
	\caption{Second derivative of the signal represented in Fig \ref{fig4}}
	\label{fig7}
\end{figure}

The differentiated signal is used to extract a feature called `Augmentation Index' (\textit{AIx}). In order to obtain \textit{AIx} the differentiated signal is again passed through a differentiator, and a butterworth low pass filter of order 2 and a cut off frequency of 250 Hz. The first order and the second order differentiated signals are individually passed through two different envelope detector in order to identify the peaks required for calculation of \textit{AIx}. Derivative estimation based technique \cite{rezk2011algebraic} has been used for envelope detection. Fig. \ref{fig7} shows the second order differentiated signal along with the two identified peaks required for calculation of \textit{AIx}.

In order to calculate the Augmentation Index (\textit{AIx}), the maximum peak voltage value obtained after the first differentiation is stored as the first peak. Time of occurrence of this first peak is also stored. The first order differentiated signal (dz/dt) is again differentiated (d\textsuperscript{2}z/dt\textsuperscript{2}) in order to identify the second change in peak voltage in the first order differentiated signal (dz/dt). The time of occurrence of this change is used to find its corresponding voltage value in the dz/dt signal. This value is saved as the second peak. Later on these two voltage values, in terms of amplitude, is used to calculate \textit{AIx}. The Augmentation Index (\textit{AIx}) obtained from the diffferentiated signal could be represented by the following equation:

\begin{equation}
		AIx = \frac{{\mbox{\textit{Amplitude of first peak}} - \mbox{\textit{Amplitude of second peak}}}}{{\mbox{\textit{Amplitude of first peak}}}}
\end{equation}

\begin{figure*}[htbp]
	\centering{\includegraphics[width=0.75\textwidth]{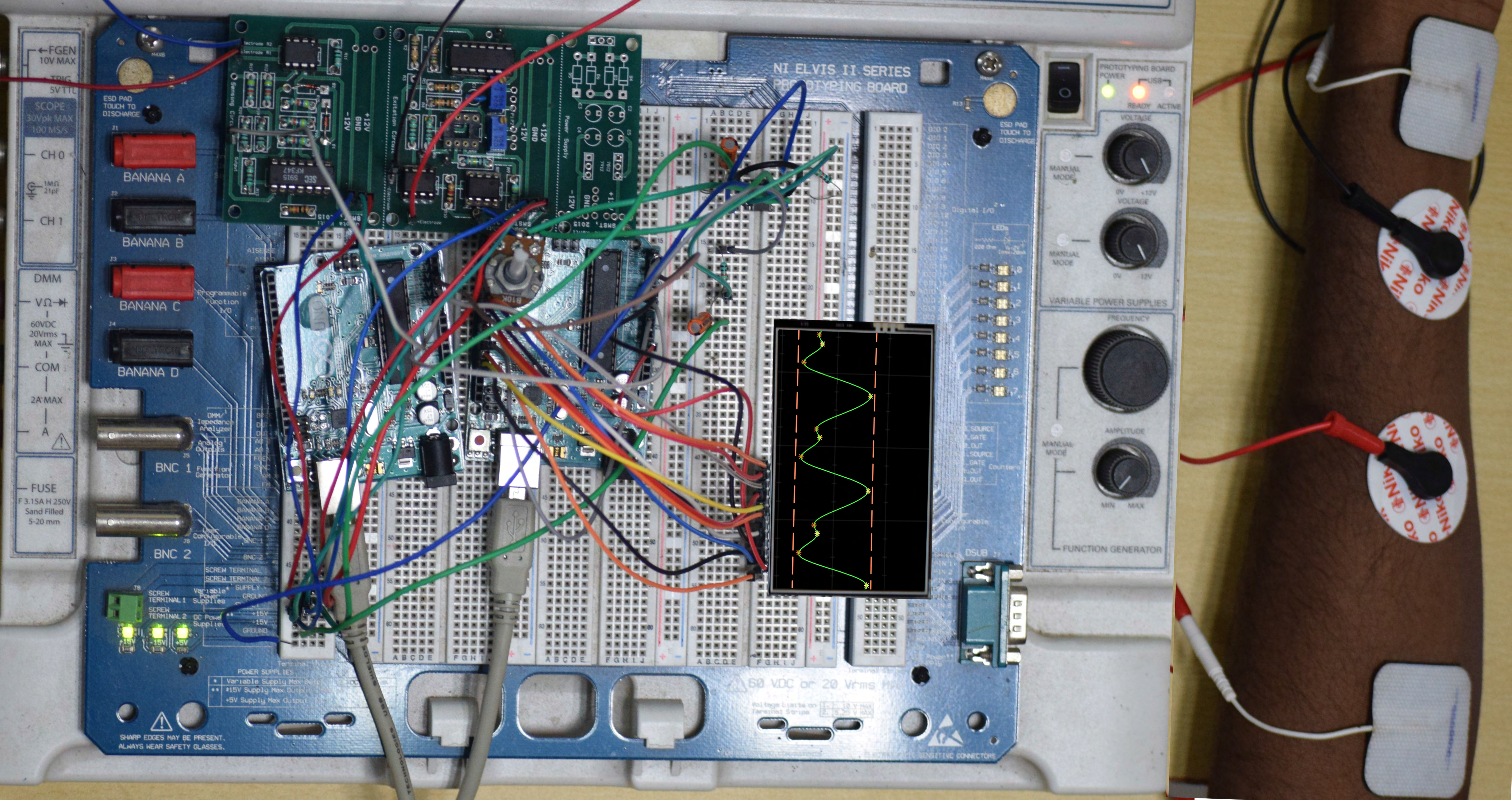}}	
	\caption{Illustrative image of the prototype developed with electrodes placed on the arm}
	\label{fig8}
\end{figure*}

All the signals are recorded at a sampling rate of 400 samples per second. Fig. \ref{fig8} shows a snapshot of the device acquiring ICG signal from a subject.

The acquired impedance cardiography (ICG) signal as seen in Fig. \ref{fig3} is very much similar to a conventional aortic pulse waveform \cite{qasem2008determination}. The extracted features are broadly categorized into four major feature groups: (a) Pressure features, (b) Time features, (c) Area features, (d) Amplitude features.The algorithm deployed in the ICG device is designed in such a way, so that it detects only the complete cardiac cycles and discards any incomplete cardiac cycle, which is the part of a cycle that misses out identification of any one of the following feature points:
\begin{itemize}
	\item Peak systolic point
	\item Dicrotic notch
	\item Peak diastolic point
	\item Baseline point
\end{itemize} 
Once the complete cardiac cycles are identified, the amplitudes of the first (P1), second (P2), third (P3) and fourth (P4) peak points are stored (where P1 \& P3 are the peak systolic points, P2 \& P4 are the peak diastolic points of two consecutive complete cardiac cycles), and the difference in amplitudes between these points are calculated. The difference values are termed as `P1-P2' and `P3-P4'. The first order and the second order differentiated ICG waveforms are used to extract augmentation index (\textit{AIx}). The method employed to extract \textit{AIx} is shown in Fig. \ref{fig5}. The filtered ICG wave, as  shown in Fig. \ref{fig4}, is used to extract the remaining feature points. The detailed features extracted from the filtered and differentiated ICG signal are shown in Table \ref{table2}. Amongst all these features, amplitude features are the major contributors for localization and detection of lesion in the coronary arteries. 
\begin{table*}[]
	\centering
	\caption{Features extracted from the ICG signal}
	\label{table2}
	\resizebox{0.75\textwidth}{!}{%
		\begin{tabular}{@{}ll@{}}
			\toprule
			Type & Description \\ \midrule
			\multicolumn{1}{c}{\multirow{5}{*}{Pressure features}} & Systolic blood pressure \\
			\multicolumn{1}{c}{} & Diastolic blood pressure \\
			\multicolumn{1}{c}{} & Peak systolic pressure \\
			\multicolumn{1}{c}{} & Peak diastolic pressure \\
			\multicolumn{1}{c}{} & Mean blood pressure \\ \midrule
			\multirow{4}{*}{Time features} & Heart rate \\  
			& Systolic duration \\  
			& Diastolic duration \\  
			& Cardiac cycle \\ \midrule
			\multirow{4}{*}{Area features} & Area under one cardiac cycle \\  
			& Systolic area \\  
			& Diastolic area \\ 
			& Area under differentiated wave \\ \midrule
			\multirow{3}{*}{Amplitude features} & Difference of first two peaks of the extracted cardiac cycle (P1-P2) \\  
			& Difference of next two peaks of the extracted cardiac cycle (P3-P4) \\  
			& Augmentation index (\textit{AIx}) \\ \bottomrule
		\end{tabular}%
	}
\end{table*}

\subsection{Implementation of artificial neural network}
\begin{figure}[htbp]
	\centering{\includegraphics[width=\columnwidth]{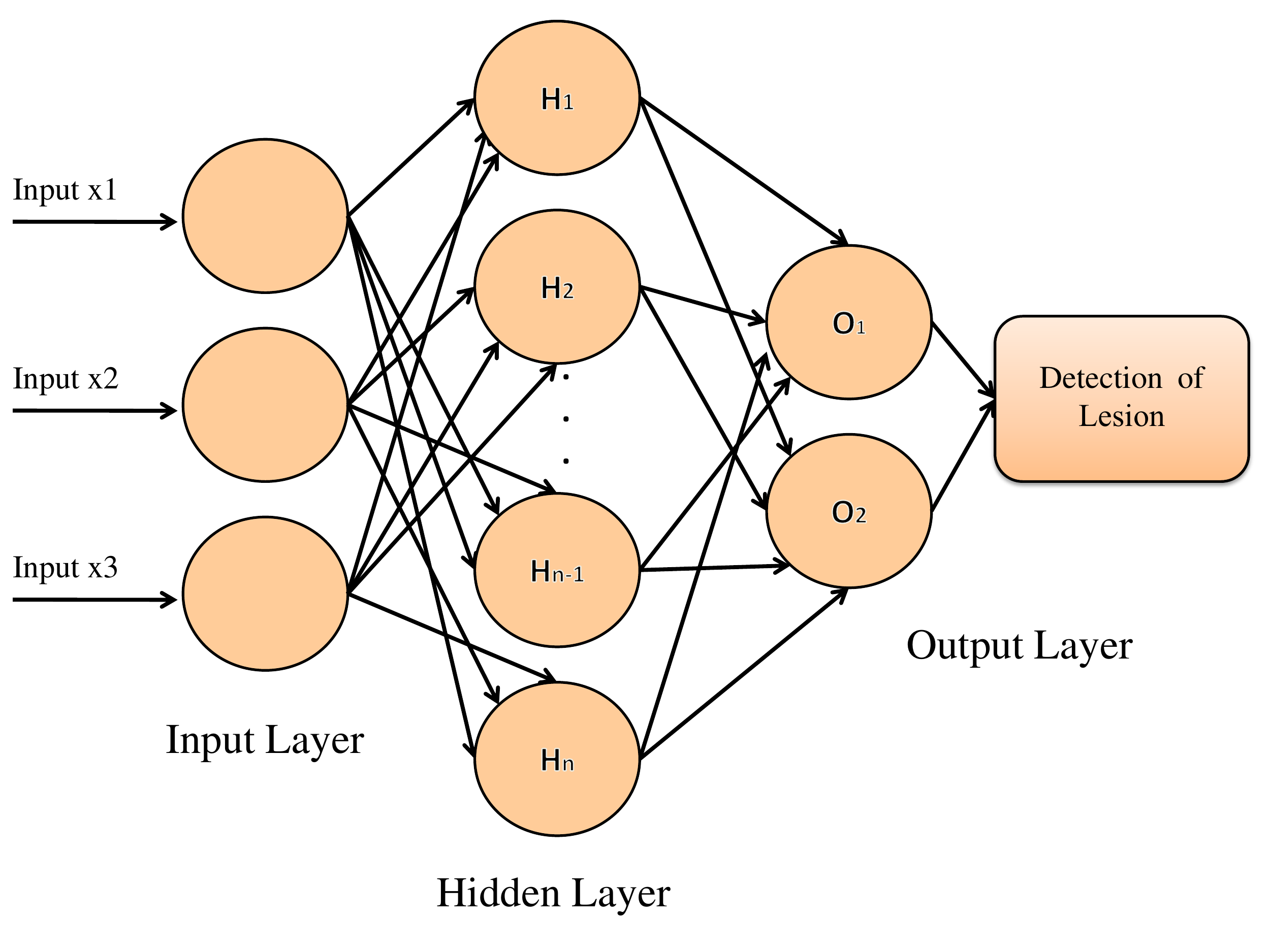}}
	\caption{Schematic view of the implemented ANN}
	\label{fig9}
\end{figure}

\begin{figure}[htbp]
	\centering{\includegraphics[width=\columnwidth]{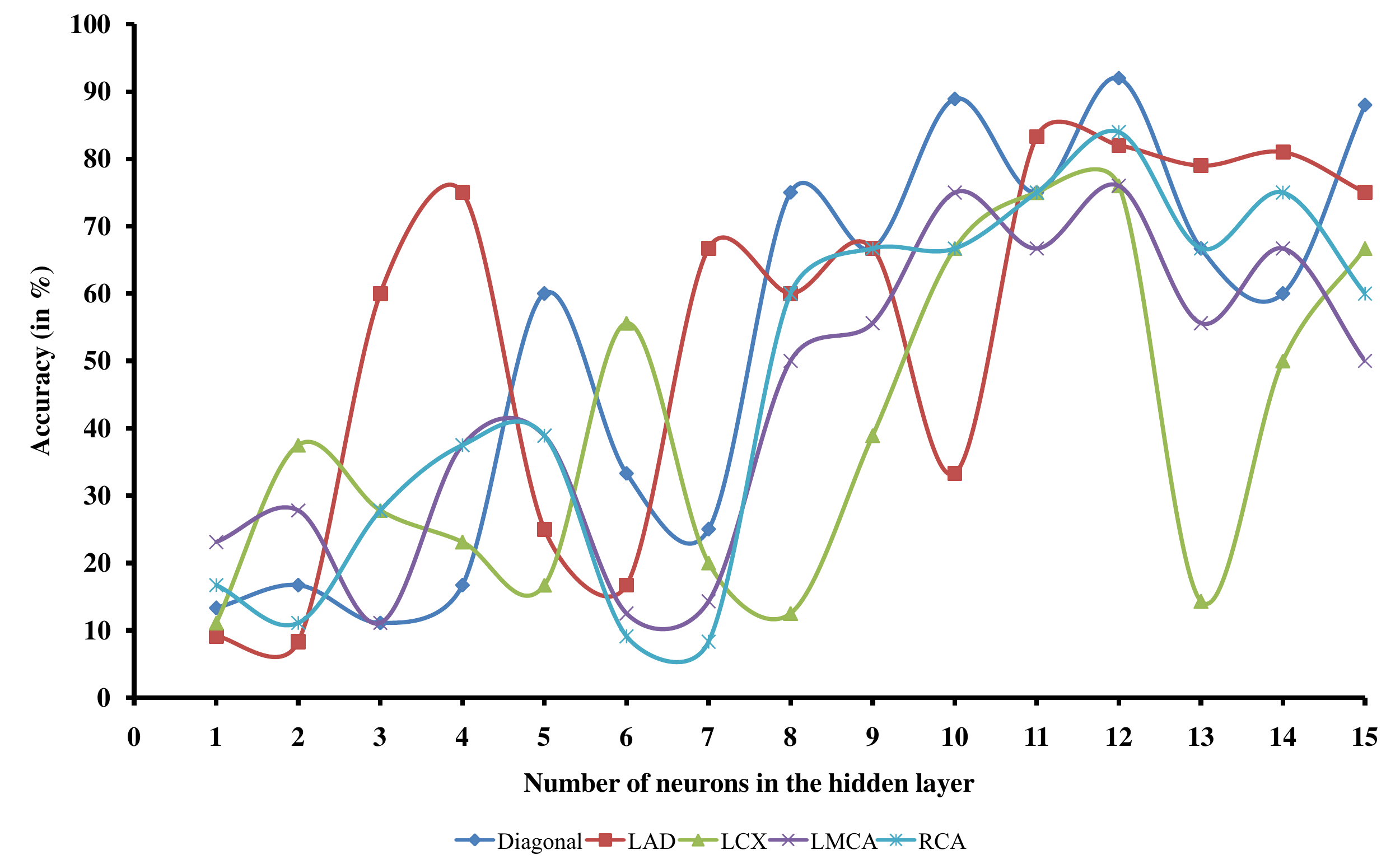}}
	\caption{Accuracy obtained for the various branches of coronary artery while varying the number of neurons in the hidden layer}
	\label{fig9_1}
\end{figure}

Detection and localization of arterial lesion is done with the aid of artificial neural network (ANN). The amplitude features of the ICG signal are fed into a three layer feed forward ANN, to determine the location and occurrence of lesion. Fig. \ref{fig9} shows the basic architecture of the implemented ANN.
The input layer consists of the amplitude features extracted from the ICG signal. The hidden layer consists of several neurons, which connects the input layer with the output layer. In order to derive an appropriate architecture of the artificial neural network, with the intention of keeping its number of parameters (weights) as small as possible, variations of the network in terms of  number of neurons in the hidden layer, is performed. The number of neurons in the hidden layer is varied from 1 to 15, and the accuracy obtained is observed. Fig. \ref{fig9_1} gives us a graphical representation of the accuracy obtained for the various branches of coronary artery. From the results obtained, it could be safely assumed that the ANN with 12 neurons in the hidden layer gives us the optimum performance, and hence the number of neurons in the hidden layer is kept as 12. The hidden layer neurons (12 neurons were considered) are defined by a logarithmic sigmoid transfer function. The outputs from the hidden and input layers are simultaneously connected and fed to the neurons of the output layer. Each neuron in the hidden layer of the implemented network can be represented by Eq. \ref{eq3}, where $\sigma$ is the sigmoid function, ${\vec w}$ is the weight vector, and ${\vec x}$ represents the input vector. The output layer identifies the occurence of lesion. For each of the branch of coronary artery, separate individual ANNs are used (i.e. one ANN for each of the coronary arteial branch). The ANN implemented is trained separately, such that its functions are generated for each of the 5 coronary arterial branch, thereby localising the occurence of lesion using all the 5 networks developed. 

\begin{equation}
		\centering
		Output = \sigma (\vec w \cdot \vec x)
		\label{eq3}
\end{equation}

where,

\begin{equation}
		\centering
		\sigma (y) = \frac{1}{{1 + {e^{ - y}}}}
		\label{eq4}
\end{equation}

In order to train and test the implemented ANN, the data collected from 50 subjects are further sub divided into training data, and testing data, based on k-fold cross validation, where k=5. 5-fold cross validation, is implemented to validate the models predicted. The entire data set is split into 5 groups , each containing equal number of data (but no repeated data). Then one of the data group is held out as a test data set and the remaining data sets are treated as training data sets. The model is then trained using the training data sets and then tested on the test data set. As a result of which, at every iteration 1/5th of the subject population (ten samples) is reserved as test data set and remaining 4/5th of the subject population (forty samples) is treated as training data set. For each iteration, the evaluation score is retained. Importantly, each observation in the data sample is assigned to an individual group and stays in that group for the duration of the procedure. This means that each sample is given the opportunity to be used in the hold out set 1 time and used to train the model k-1 times. At the end the performance of the model is predicted based on the sample of model evaluation scores. The training data set is also employed to optimise the weight function, which links the input layer with the hidden layer. The actual occurence of lesion are ascertained by performing invasive coronary angiogram. 

\subsection{Coronary angiogram}
\begin{figure}[htbp]
	\centering{\includegraphics[width=\columnwidth]{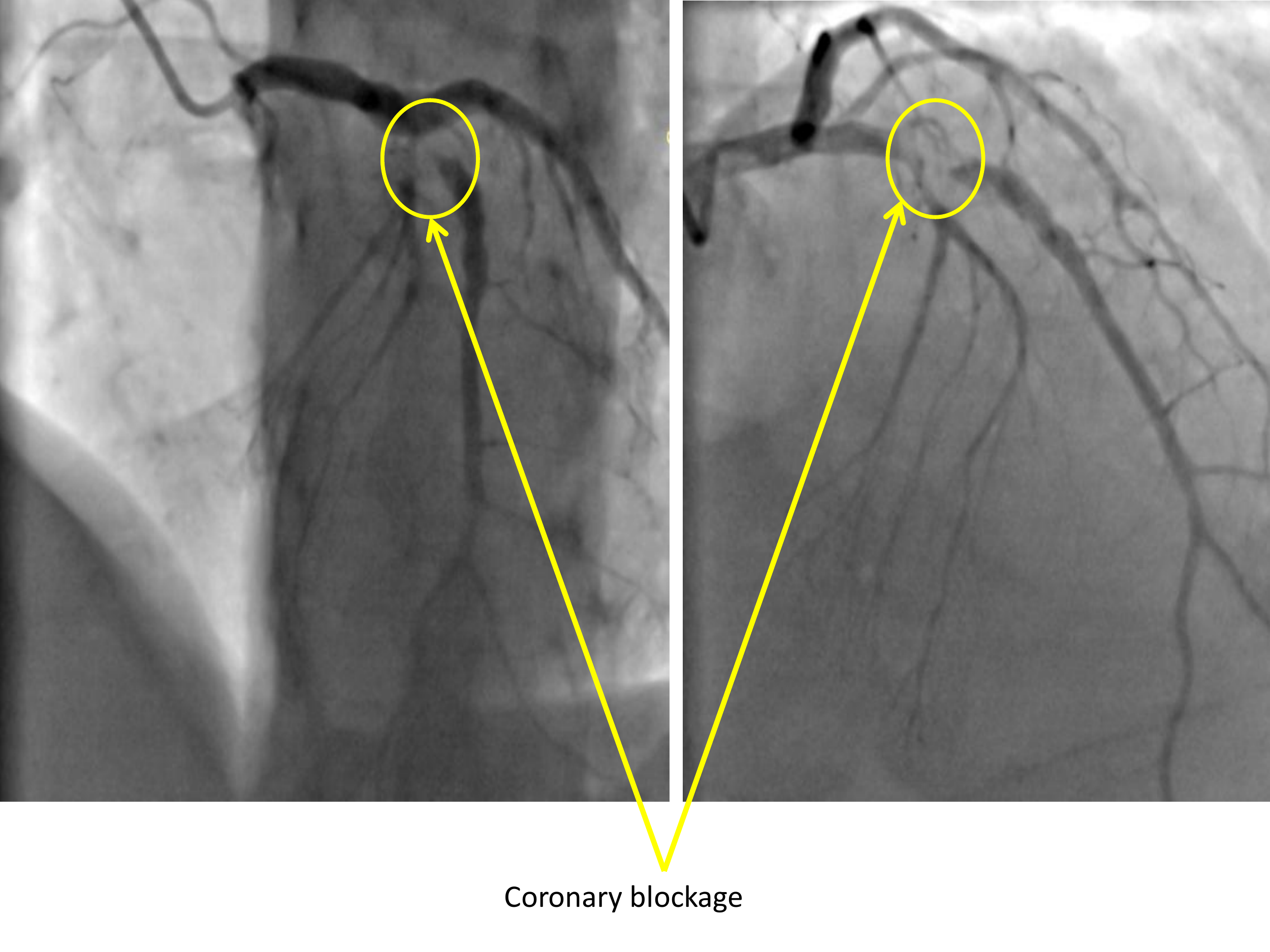}}	
	\vspace{-20pt}
	\caption{Coronary angiogram depicting blockage in Left Anterior Descending(LAD) artery}
	\label{fig10}
\end{figure}

Coronary angiogram (CAG) is an invasive procedure, performed to ascertain the occurence of lesion in coronary arteries \cite{galbraith1978coronary}. All the 50 subjects recruited in the study, underwent coronary angiogram at ``Medical College and Hospital, Kolkata", under expert supervision. The results obtained from coronary angiogram were further analysed to ascertain the location and occurence of lesion. The subject's coronary angiogram were mainly used to identify coronary blockages in the following branches of artery:
\begin{itemize}
	\item Left main coronary artery (LMCA)
	\item Left anterior descending artery (LAD)
	\item Diagonal branch
	\item Left circumflex artery (LCX)	
	\item Right coronary artery (RCA)
\end{itemize}
CAG is considered to be the `gold standard' when it comes to detection of lesion.The device used for performing coronary angiogram is ``Siemens\textsuperscript{TM} Axiom Artis Zee (floor)", with a power rating of 100 kW at 100 kV. The power of the X-ray generator is 100 kW, with a penetration depth of 92 cm. The device is capable of making right anterior oblique (rao) projection of $\pm$ 130 and left anterior oblique (lao) projection of 120 at 25/second. The focal spot area varies from 0.3 mm to 1.0 mm. Fig. \ref{fig10} shows two different views of coronary blockage of Left Anterior Descending(LAD) artery, identified by CAG.

\section{Results and Discussion}

\begin{figure*}[htbp]
	\begin{subfigure}[b]{0.33\textwidth}
		\includegraphics[width=\linewidth]{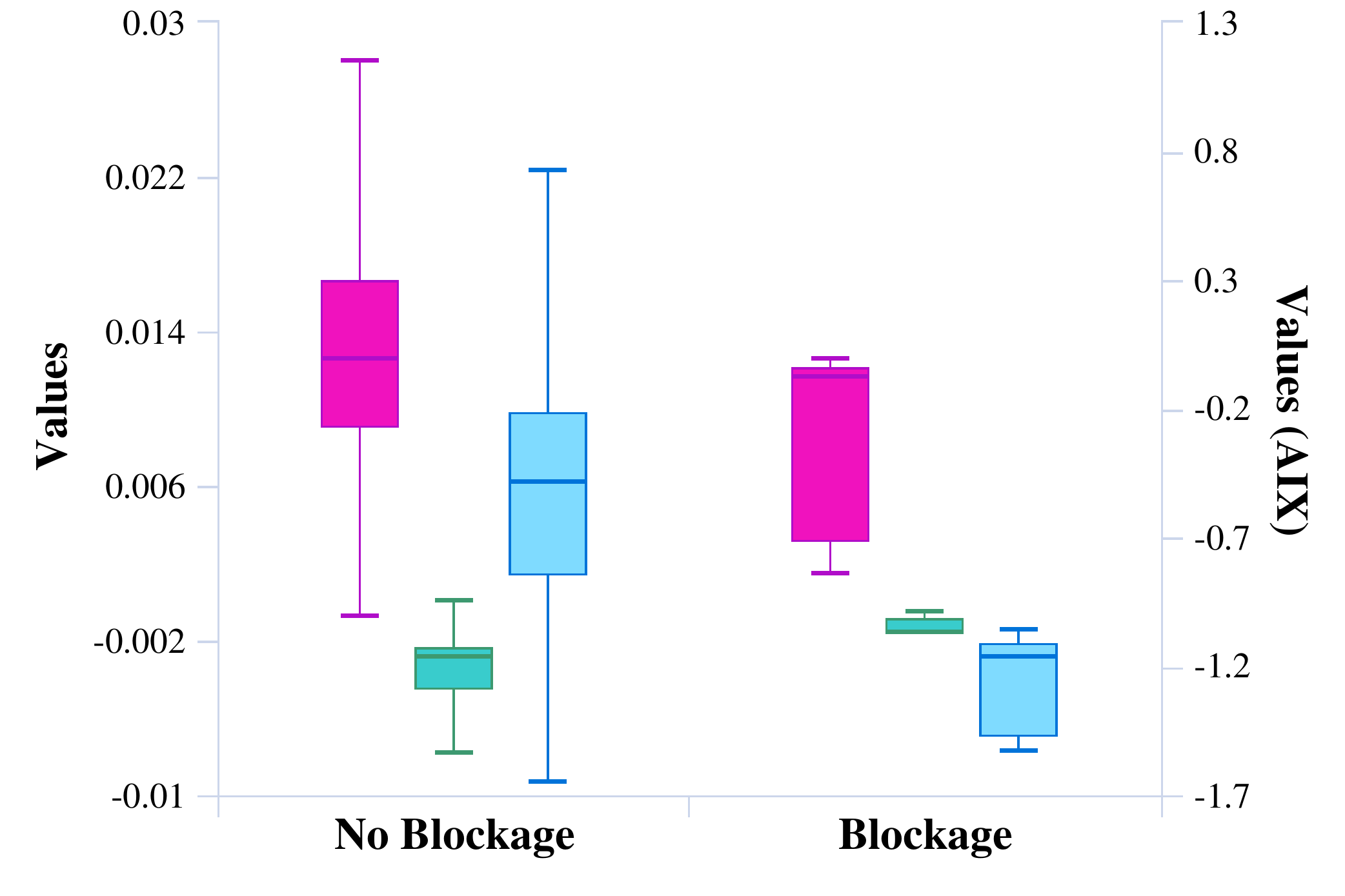}
		\caption{Diagonal}
		\label{fig11a}
	\end{subfigure}%
	\begin{subfigure}[b]{0.33\textwidth}
		\includegraphics[width=\linewidth]{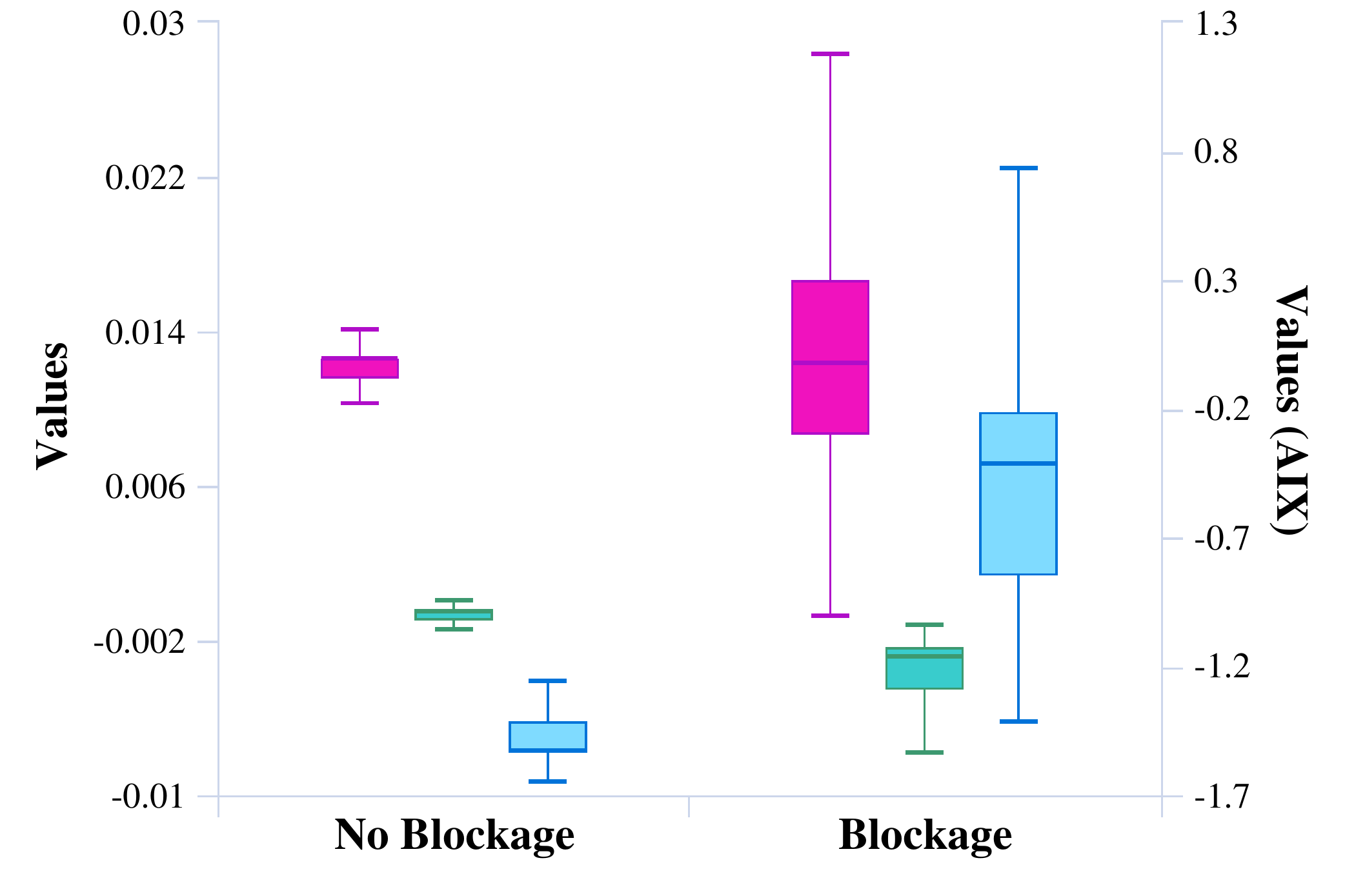}
		\caption{LAD}
		\label{fig11b}
	\end{subfigure}%
	\begin{subfigure}[b]{0.33\textwidth}
		\includegraphics[width=\linewidth]{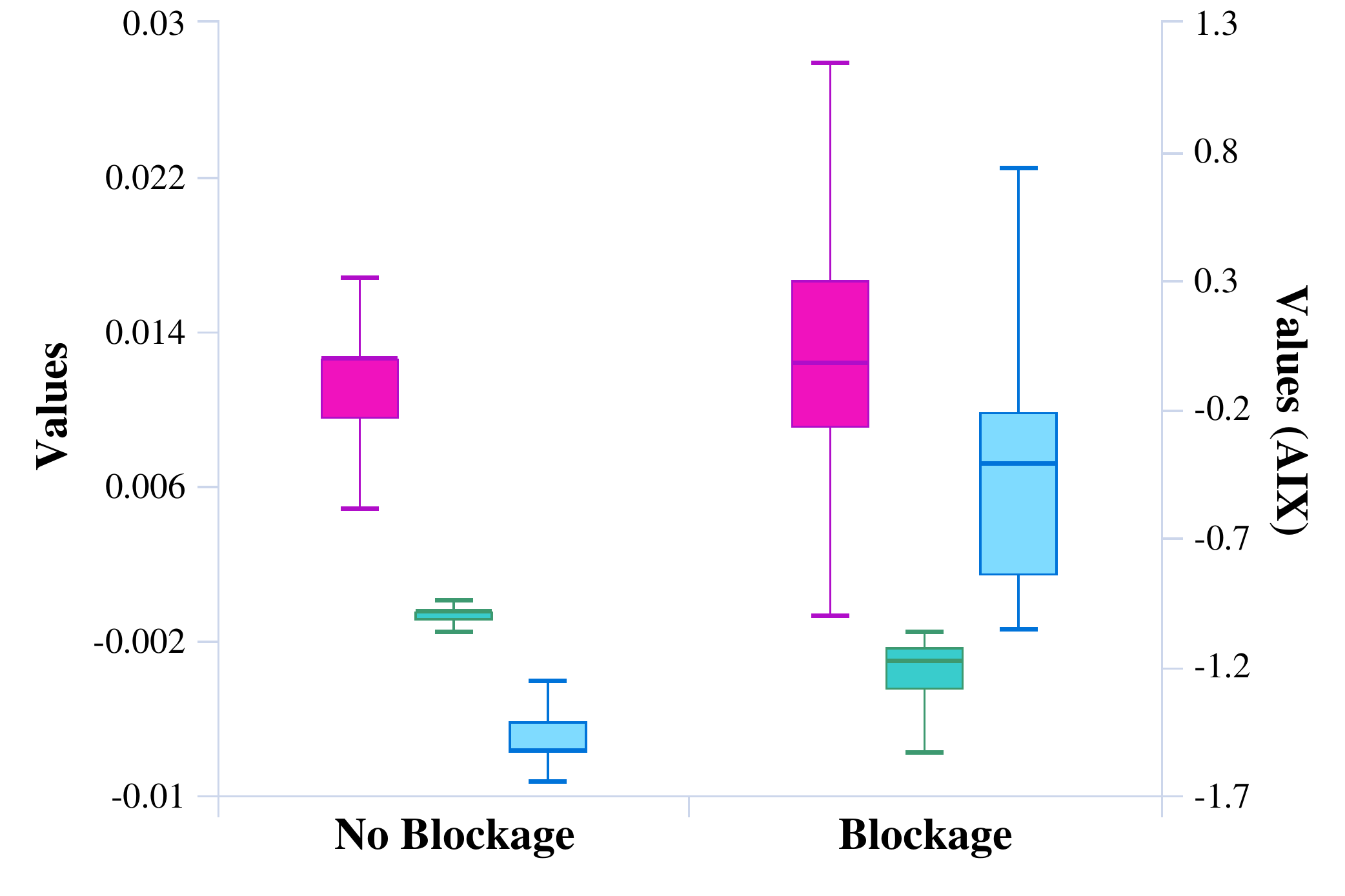}
		\caption{LCX}
		\label{fig11c}
	\end{subfigure}%
	\\[1ex]
	\centering
	\begin{subfigure}[b]{0.33\textwidth}
		\includegraphics[width=\linewidth]{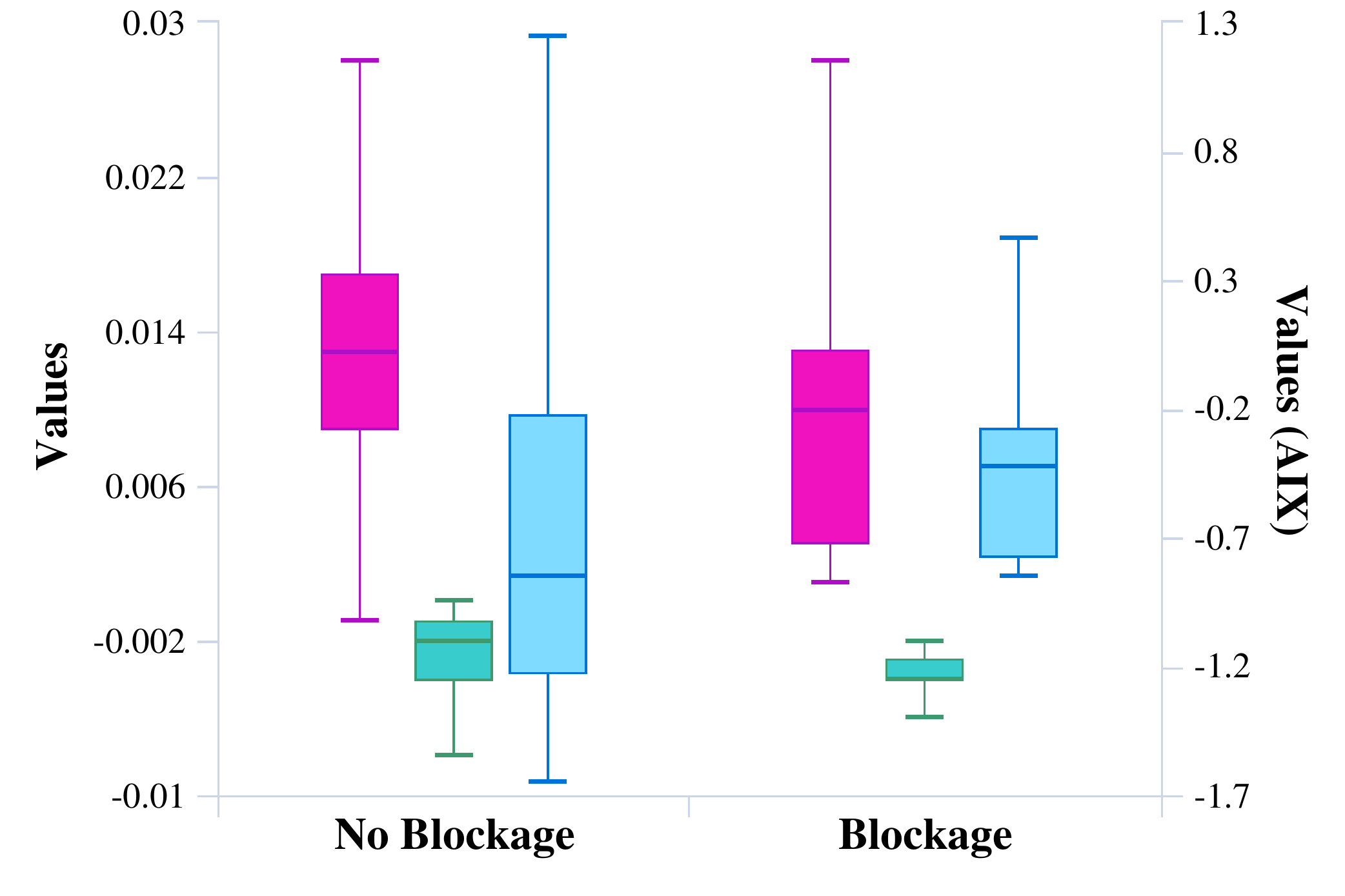}
		\caption{LMCA}
		\label{fig11d}
	\end{subfigure}%
	\begin{subfigure}[b]{0.33\textwidth}
		\includegraphics[width=\linewidth]{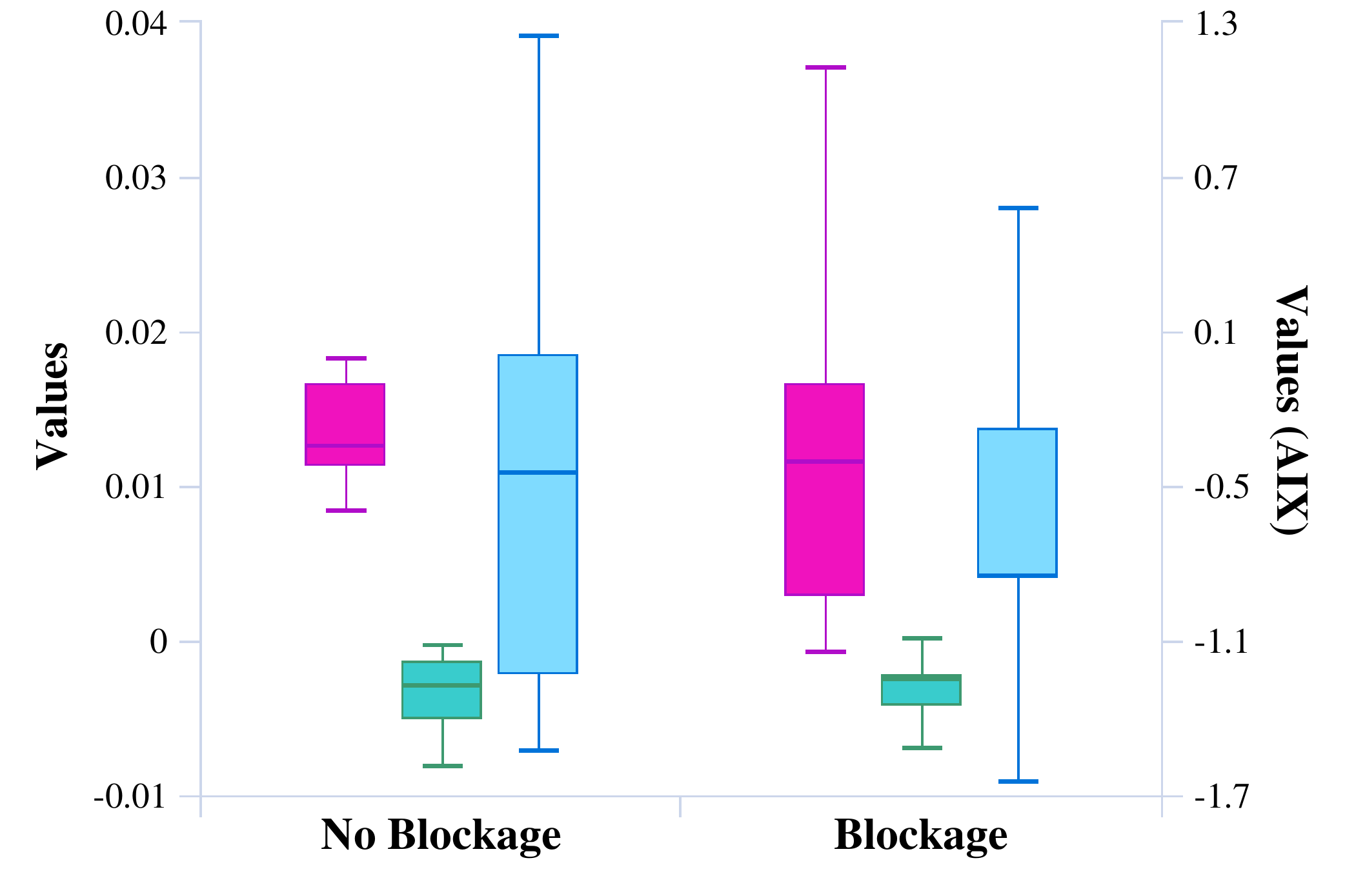}
		\caption{RCA}
		\label{fig11e}
	\end{subfigure}%
	\\[1ex]
	\centering
	\begin{subfigure}[b]{0.15\textwidth}
		\includegraphics[width=\linewidth]{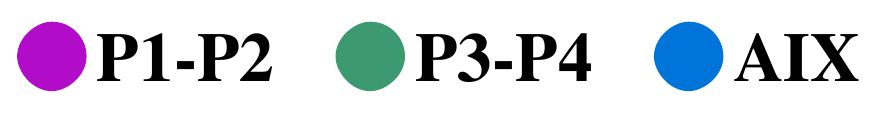}
	\end{subfigure}
	\caption{Boxplots of parameters with \textit{AIx} plotted on secondary axis}
	\label{fig11}
\end{figure*}

The impedance cardiography set up as shown in Fig. \ref{fig8} is used to acquire ICG signals from the recruited subjects. Coronary angiogram of respective subjects is also performed. The ICG signal acquired is also affected by both cardiac and vascular properties \cite{avolio2013arterial}, \cite{weber2001measurement}. Since the heart is affected by coronary arterial blocakages, the shape of the acquired arterial pulse (in the form of ICG signal) also varies. The variations in the shape of the ICG signal is reflected in the amplitude parameters (at the peak systolic [P1 \& P3] and peak diastolic [P2 \& P4] points) extracted from the ICG signal. The ICG signal acquired is filtered and differentiated to extract certain features for further analysis. The amplitude features extracted, as shown in Table \ref{table2}, are used for prediction of blockages in the coronary arteries. The first and foremost method applied is to identify blockages based on simple linear satistical analysis. The mean and standard deviation of `P1-P2', `P3-P4' and `\textit{AIx}' are calculated, but as shown in Fig. \ref{fig11}, the standard deviation was found to be very high. 
Correlation between these parameters and blockages in various branches of coronary artery is also studied. The correlation values obtained are shown in Table \ref{table3}. It is seen that the parameters are not significantly correlated. As a result of these it can be said that linear statistical analysis is not able to provide us with significant results. The same parameters are then studied using artificial neural network (ANN).
\begin{table}[htbp]
	\centering
	\caption{Correlation between extracted features and localised area of lesion}
	\label{table3}
	\resizebox{\columnwidth}{!}{%
		\begin{tabular}{l|c|c|c|c|c|}
			\cline{2-6}
			& \multicolumn{1}{l|}{\textit{\textbf{Diagonal}}} & \multicolumn{1}{l|}{\textit{\textbf{LAD}}} & \multicolumn{1}{l|}{\textit{\textbf{LCX}}} & \multicolumn{1}{l|}{\textit{\textbf{LMCA}}} & \multicolumn{1}{l|}{\textit{\textbf{RCA}}} \\ \hline
			\multicolumn{1}{|l|}{\textit{\textbf{P1-P2}}} & 0 & -0.027 & -0.028 & -0.058 & -0.723 \\ \hline
			\multicolumn{1}{|l|}{\textit{\textbf{P3-P4}}} & 0 & -0.64 & -0.51 & -0.619 & -0.183 \\ \hline
			\multicolumn{1}{|l|}{\textit{\textbf{\textit{AIx}}}} & 0 & 0.42 & 0.63 & 0.259 & 0.204 \\ \hline
		\end{tabular}%
	}
\end{table}
A three layered feed forward ANN, with twelve neurons in the hidden layer is implemented to establish a relation between the extracted amplitude parameters and arterial blockages. Results obtained from CAG performed on the subjects, serve as the gold standard data for coronary blockages. The ANN is trained to identify any blockages and the location of the blockage. The network is trained with data for blockages from the following coronary arteries:
\begin{itemize}
	\item Left main coronary artery (LMCA)
	\item Left anterior descending artery (LAD)
	\item Diagonal branch
	\item Left circumflex artery (LCX)	
	\item Right coronary artery (RCA)
\end{itemize}
The amplitude parameters extracted from the ICG signal acts as the input to the network. The network is trained separately for all the branches of coronary artery, as specified above. K-fold cross validation with k=5, is implemented in order to test and train the implemented artificial neural network and verify the results obtained. The obtained predictions from the trained network is validated against data available from coronary angiogram. Fig. \ref{fig12}, \ref{fig13}, \ref{fig14}, \ref{fig15}, and \ref{fig16} shows the confusion matrix obtained by feeding the input parameters into the trained network. 
\begin{figure}[htbp]
	\centering
	\resizebox{\columnwidth}{!}{%
		\begin{tabular}{cl|l|l|l|}
			\cline{3-5}
			\multicolumn{1}{l}{} &  & \multicolumn{3}{c|}{\textit{\textbf{Actual}}} \\ \cline{2-5} 
			\multicolumn{1}{l|}{} & \multicolumn{1}{c|}{n=50} & Blockage & No Blockage & \textbf{Total} \\ \hline
			\multicolumn{1}{|c|}{\multirow{3}{*}{\textit{\textbf{Predicted}}}} & Blockage & 5 & 2 & \textbf{7} \\ \cline{2-5} 
			\multicolumn{1}{|c|}{} & No Blockage & 2 & 41 & \textbf{43} \\ \cline{2-5} 
			\multicolumn{1}{|c|}{} & \textbf{Total} & \textbf{7} & \textbf{43} & \textbf{50} \\ \hline
		\end{tabular}%
	}
	\caption{Confusion matrix for localization of the diagonal branch of artery}
	\label{fig12}
\end{figure}
\begin{figure}[htbp]
	\centering
	\resizebox{\columnwidth}{!}{%
		\begin{tabular}{cl|l|l|l|}
			\cline{3-5}
			\multicolumn{1}{l}{} &  & \multicolumn{3}{c|}{\textit{\textbf{Actual}}} \\ \cline{2-5} 
			\multicolumn{1}{l|}{} & \multicolumn{1}{c|}{n=50} & Blockage & No Blockage & \textbf{Total} \\ \hline
			\multicolumn{1}{|c|}{\multirow{3}{*}{\textit{\textbf{Predicted}}}} & Blockage & 39 & 9 & \textbf{48} \\ \cline{2-5} 
			\multicolumn{1}{|c|}{} & No Blockage & 0 & 2 & \textbf{2} \\ \cline{2-5} 
			\multicolumn{1}{|c|}{} & \textbf{Total} & \textbf{39} & \textbf{11} & \textbf{50} \\ \hline
		\end{tabular}%
	}
	\caption{Confusion matrix for localization of the LAD branch of artery}
	\label{fig13}
\end{figure}
\begin{figure}[htbp]
	\centering
	\resizebox{\columnwidth}{!}{%
		\begin{tabular}{cl|l|l|l|}
			\cline{3-5}
			\multicolumn{1}{l}{} &  & \multicolumn{3}{c|}{\textit{\textbf{Actual}}} \\ \cline{2-5} 
			\multicolumn{1}{l|}{} & \multicolumn{1}{c|}{n=50} & Blockage & No Blockage & \textbf{Total} \\ \hline
			\multicolumn{1}{|c|}{\multirow{3}{*}{\textit{\textbf{Predicted}}}} & Blockage & 26 & 0 & \textbf{26} \\ \cline{2-5} 
			\multicolumn{1}{|c|}{} & No Blockage & 12 & 12 & \textbf{24} \\ \cline{2-5} 
			\multicolumn{1}{|c|}{} & \textbf{Total} & \textbf{38} & \textbf{12} & \textbf{50} \\ \hline
		\end{tabular}%
	}	
	\caption{Confusion matrix for localization of the LCX branch of artery}
	\label{fig14}
\end{figure}
\begin{figure}[htbp]
	\centering
	\resizebox{\columnwidth}{!}{%
		\begin{tabular}{cl|l|l|l|}
			\cline{3-5}
			\multicolumn{1}{l}{} &  & \multicolumn{3}{c|}{\textit{\textbf{Actual}}} \\ \cline{2-5} 
			\multicolumn{1}{l|}{} & \multicolumn{1}{c|}{n=50} & Blockage & No Blockage & \textbf{Total} \\ \hline
			\multicolumn{1}{|c|}{\multirow{3}{*}{\textit{\textbf{Predicted}}}} & Blockage & 2 & 4 & \textbf{6} \\ \cline{2-5} 
			\multicolumn{1}{|c|}{} & No Blockage & 8 & 36 & \textbf{44} \\ \cline{2-5} 
			\multicolumn{1}{|c|}{} & \textbf{Total} & \textbf{10} & \textbf{40} & \textbf{50} \\ \hline
		\end{tabular}%
	}
	\caption{Confusion matrix for localization of the LMCA branch of artery}
	\label{fig15}
\end{figure}
\begin{figure}[htbp]
	\centering
	\resizebox{\columnwidth}{!}{%
		\begin{tabular}{cl|l|l|l|}
			\cline{3-5}
			\multicolumn{1}{l}{} &  & \multicolumn{3}{c|}{\textit{\textbf{Actual}}} \\ \cline{2-5} 
			\multicolumn{1}{l|}{} & \multicolumn{1}{c|}{n=50} & Blockage & No Blockage & \textbf{Total} \\ \hline
			\multicolumn{1}{|c|}{\multirow{3}{*}{\textit{\textbf{Predicted}}}} & Blockage & 28 & 6 & \textbf{34} \\ \cline{2-5} 
			\multicolumn{1}{|c|}{} & No Blockage & 2 & 14 & \textbf{16} \\ \cline{2-5} 
			\multicolumn{1}{|c|}{} & \textbf{Total} & \textbf{30} & \textbf{20} & \textbf{50} \\ \hline
		\end{tabular}%
	}
	\caption{Confusion matrix for localization of the RCA branch of artery}
	\label{fig16}
\end{figure}
The obtained confusion matrix is further used to calculate various performance parameters of the trained network, using the following equations \cite{glaros1988understanding}.
\begin{equation}
		Prevalance = \frac{{\sum {tp} + \sum {fn} }}{n}
\end{equation}

\begin{equation}
		Accuracy = \frac{{\sum {tp + \sum {tn} } }}{n}
\end{equation}

\begin{equation}
		Precision = \frac{{\sum {tp} }}{{\sum {tp} + \sum {fn} }}
\end{equation}

\begin{equation}
		Sensitivity = \frac{{\sum {tp} }}{{\sum {tp} + \sum {fn} }}
\end{equation}

\begin{equation}
		Specificity = \frac{{\sum {tn} }}{{\sum {fp} + \sum {tn} }}
\end{equation}

\begin{equation}
		F - score = \frac{2}{{\frac{1}{{Sensitivity}} + \frac{1}{{Precision}}}}
\end{equation}

The attributes used to calculate the above parameters for detection and localization of lesion are as follows:
\begin{itemize}
	\item \textit{tp}- true positive, i.e. The subject has  blockage in the coronary artery, and it is classified as blockage in that particular coronary artery.
	\item \textit{tn}- true negative, i.e. The subject has no blockage in a particular coronary artery, and it is classified as no blockage in that particular coronary artery. 
	\item \textit{fp}- false positive, i.e. The subject has no blockage in a particular coronary artery, and it is classified as blockage in that particular coronary artery.
	\item \textit{fn}- false negative, i.e. The subject has blockage in a particular coronary artery, and it is classified as no blockage in that particular coronary artery.
	\item \textit{n}- Total number of subjects recruited in the study
\end{itemize}

In a previous work by, Paradkar \textit{et al.} \cite{paradkar2017coronary}, PPG was used for detection of coronary arterial blockage. However, the location of occurrence of blockage was not reported. The sensitivity and specificity values reported by Paradkar \textit{et al.} \cite{paradkar2017coronary}, were 85 \% and 78 \%, respectively. In this work, we are not only able to diagnose the occurrence of coronary arterial blockage (lesion), but are also able to detect the location of the blockage.

The trained network is able to correctly identify 46 cases out of 50 in case of Diagonal (as seen in Fig. \ref{fig12}), 41 out of 50 in case of LAD (as seen in Fig. \ref{fig13}),  38 out of 50 in case of LCX (as seen in Fig. \ref{fig14}), 38 out of 50 in case of LMCA (as seen in Fig. \ref{fig15}) and 42 out of 50 in case of RCA (as seen in Fig. \ref{fig16}) branch of coronary arteries. The accuracy of the predicted localised area of lesion is 92\%, 82\%, 76\%, 76\%, and 84\% in case of Diagonal, LAD, LCX, LMCA, RCA respectively. Precision, Sensitivity, Specificity, Prevelance and F-score of the trained network is also calculated and is shown in Table \ref{table4}. 
\begin{table*}[]
	\centering
	\caption{Performance parameters of the trained network for localization of occurence of lesion}
	\label{table4}
	\resizebox{\textwidth}{!}{%
		\begin{tabular}{l|c|c|c|c|c|c|}
			\cline{2-7}
			& \multicolumn{1}{l|}{\textit{\textbf{Prevalance (\%)}}} & \multicolumn{1}{l|}{\textit{\textbf{Accuracy (\%)}}} & \multicolumn{1}{l|}{\textit{\textbf{Precision (\%)}}} & \multicolumn{1}{l|}{\textit{\textbf{Sensitivity (\%)}}} & \multicolumn{1}{l|}{\textit{\textbf{Specificity (\%)}}} & \multicolumn{1}{l|}{\textit{\textbf{F-Score}}} \\ \hline
			\multicolumn{1}{|l|}{\textit{\textbf{Diagonal}}} & 14 & 92 & 71.42 & 71.42 & 95.34 & 0.71 \\ \hline
			\multicolumn{1}{|l|}{\textit{\textbf{LAD}}} & 78 & 82 & 81.25 & 100 & 18.18 & 0.89 \\ \hline
			\multicolumn{1}{|l|}{\textit{\textbf{LCX}}} & 76 & 76 & 100 & 68.42 & 100 & 0.82 \\ \hline
			\multicolumn{1}{|l|}{\textit{\textbf{LMCA}}} & 20 & 76 & 33.33 & 20 & 90 & 0.25 \\ \hline
			\multicolumn{1}{|l|}{\textit{\textbf{RCA}}} & 60 & 84 & 82.35 & 93.33 & 70 & 0.87 \\ \hline
		\end{tabular}%
	}
\end{table*}

\section{Conclusion}
The primary objective of this study, is to develop a novel technique for  non-invasive detection and localization of arterial lesion. The current methods employed for detection of lesion are invasive and costly procedure, which also involves a certain amount of risk due to surgical complications. The proposed device along with its implemented algorithms is not only able to identify blockages, but also is able to specify the location of arterial blockage with high amount of accuracy. The obtained results are quite promising and thereby increases the scope of using ICG for detection and localization of arterial blockages. The proposed device if used on a large scale would bring down the cost associated with coronary angiogram and would also provide clinicians with means to identify coronary blockages without using any surgical set up. The proposed device could also be used by any person for self diagnosis of arterial lesion. We expect that the proposed methodology  would lead to the use of Impedance Cardiography in real life clinical practices.

\section*{Acknowledgment}
The work described in this paper was funded by MHRD, Department of Higher Education, New Delhi, INDIA, F. NO. 4-23/2014-TS.I, Dt. 14-02-2014, (project code: LYA). The authors declare that a patent has been filed for the device used in this work, vide application number KOL/201831001822.

\bibliographystyle{IEEEtran}
\bibliography{references}

\begin{thebibliography}{10}
\providecommand{\url}[1]{#1}
\csname url@samestyle\endcsname
\providecommand{\newblock}{\relax}
\providecommand{\bibinfo}[2]{#2}
\providecommand{\BIBentrySTDinterwordspacing}{\spaceskip=0pt\relax}
\providecommand{\BIBentryALTinterwordstretchfactor}{4}
\providecommand{\BIBentryALTinterwordspacing}{\spaceskip=\fontdimen2\font plus
\BIBentryALTinterwordstretchfactor\fontdimen3\font minus
  \fontdimen4\font\relax}
\providecommand{\BIBforeignlanguage}[2]{{%
\expandafter\ifx\csname l@#1\endcsname\relax
\typeout{** WARNING: IEEEtran.bst: No hyphenation pattern has been}%
\typeout{** loaded for the language `#1'. Using the pattern for}%
\typeout{** the default language instead.}%
\else
\language=\csname l@#1\endcsname
\fi
#2}}
\providecommand{\BIBdecl}{\relax}
\BIBdecl

\bibitem{who2017}
\BIBentryALTinterwordspacing
W.~H. Organization. Cardiovascular diseases (cvds). [Online]. Available:
  \url{http://www.who.int/mediacentre/factsheets/fs317/en/}
\BIBentrySTDinterwordspacing

\bibitem{benjamin2017heart}
E.~J. Benjamin, M.~J. Blaha, S.~E. Chiuve, M.~Cushman, S.~R. Das, R.~Deo, S.~D.
  de~Ferranti, J.~Floyd, M.~Fornage, C.~Gillespie \emph{et~al.}, ``Heart
  disease and stroke statistics—2017 update: a report from the american heart
  association,'' \emph{Circulation}, vol. 135, no.~10, pp. e146--e603, 2017.

\bibitem{bigatello2002hemodynamic}
L.~Bigatello and E.~George, ``Hemodynamic monitoring.'' \emph{Minerva
  anestesiologica}, vol.~68, no.~4, pp. 219--225, 2002.

\bibitem{liu2015arteries}
C.~Liu, D.~Zheng, and A.~Murray, ``Arteries stiffen with age, but can retain an
  ability to become more elastic with applied external cuff pressure,''
  \emph{Medicine}, vol.~94, no.~41, 2015.

\bibitem{townsend2015recommendations}
R.~R. Townsend, I.~B. Wilkinson, E.~L. Schiffrin, A.~P. Avolio, J.~A. Chirinos,
  J.~R. Cockcroft, K.~S. Heffernan, E.~G. Lakatta, C.~M. McEniery, G.~F.
  Mitchell \emph{et~al.}, ``Recommendations for improving and standardizing
  vascular research on arterial stiffness,'' \emph{Hypertension}, vol.~66,
  no.~3, pp. 698--722, 2015.

\bibitem{gao2014improved}
M.~Gao, G.~Zhang, N.~B. Olivier, and R.~Mukkamala, ``Improved pulse wave
  velocity estimation using an arterial tube-load model,'' \emph{IEEE
  Transactions on Biomedical Engineering}, vol.~61, no.~3, pp. 848--858, 2014.

\bibitem{zhang2011radial}
Y.-L. Zhang, Y.-Y. Zheng, Z.-C. Ma, and Y.-N. Sun, ``Radial pulse transit time
  is an index of arterial stiffness,'' \emph{Hypertension Research}, vol.~34,
  no.~7, pp. 884--887, 2011.

\bibitem{goswami2011relevance}
D.~Goswami, B.~Chatterjee, S.~Ray, K.~Chaudhuri, and J.~Mukherjee, ``On the
  relevance of a ppg based two pulse synthesis model for screening against
  coronary artery diseases,'' \emph{Artery Research}, vol.~5, no.~4, p. 161,
  2011.

\bibitem{vardoulis2013validation}
O.~Vardoulis, T.~G. Papaioannou, and N.~Stergiopulos, ``Validation of a novel
  and existing algorithms for the estimation of pulse transit time: advancing
  the accuracy in pulse wave velocity measurement,'' \emph{American Journal of
  Physiology-Heart and Circulatory Physiology}, vol. 304, no.~11, pp.
  H1558--H1567, 2013.

\bibitem{papaioannou2014validation}
T.~G. Papaioannou, O.~Vardoulis, and N.~Stergiopulos, ``Validation of
  algorithms for the estimation of pulse transit time: where do we stand
  today?'' \emph{Annals of biomedical engineering}, vol.~42, no.~6, p. 1143,
  2014.

\bibitem{ding2016continuous}
X.-R. Ding, Y.-T. Zhang, J.~Liu, W.-X. Dai, and H.~K. Tsang, ``Continuous
  cuffless blood pressure estimation using pulse transit time and
  photoplethysmogram intensity ratio,'' \emph{IEEE Transactions on Biomedical
  Engineering}, vol.~63, no.~5, pp. 964--972, 2016.

\bibitem{mukkamala2015toward}
R.~Mukkamala, J.-O. Hahn, O.~T. Inan, L.~K. Mestha, C.-S. Kim, H.~T{\"o}reyin,
  and S.~Kyal, ``Toward ubiquitous blood pressure monitoring via pulse transit
  time: theory and practice,'' \emph{IEEE Transactions on Biomedical
  Engineering}, vol.~62, no.~8, pp. 1879--1901, 2015.

\bibitem{westerhof2008individualization}
B.~E. Westerhof, I.~Guelen, W.~J. Stok, H.~A. Lasance, C.~A. Ascoop, K.~H.
  Wesseling, N.~Westerhof, W.~J.~W. Bos, N.~Stergiopulos, and J.~A. Spaan,
  ``Individualization of transfer function in estimation of central aortic
  pressure from the peripheral pulse is not required in patients at rest,''
  \emph{Journal of applied physiology}, vol. 105, no.~6, pp. 1858--1863, 2008.

\bibitem{swamy2009adaptive}
G.~Swamy, D.~Xu, N.~B. Olivier, and R.~Mukkamala, ``An adaptive transfer
  function for deriving the aortic pressure waveform from a peripheral artery
  pressure waveform,'' \emph{American Journal of Physiology-Heart and
  Circulatory Physiology}, vol. 297, no.~5, pp. H1956--H1963, 2009.

\bibitem{hahn2012subject}
J.-O. Hahn, A.~T. Reisner, F.~A. Jaffer, and H.~H. Asada, ``Subject-specific
  estimation of central aortic blood pressure using an individualized transfer
  function: a preliminary feasibility study,'' \emph{IEEE Transactions on
  Information Technology in Biomedicine}, vol.~16, no.~2, pp. 212--220, 2012.

\bibitem{gao2016simple}
M.~Gao, W.~C. Rose, B.~Fetics, D.~A. Kass, C.-H. Chen, and R.~Mukkamala, ``A
  simple adaptive transfer function for deriving the central blood pressure
  waveform from a radial blood pressure waveform,'' \emph{Scientific reports},
  vol.~6, p. 33230, 2016.

\bibitem{goswami2010new}
D.~Goswami, K.~Chaudhuri, and J.~Mukherjee, ``A new two-pulse synthesis model
  for digital volume pulse signal analysis,'' \emph{Cardiovascular
  Engineering}, vol.~10, no.~3, pp. 109--117, 2010.

\bibitem{deb2007cuff}
S.~Deb, C.~Nanda, D.~Goswami, J.~Mukhopadhyay, and S.~Chakrabarti, ``Cuff-less
  estimation of blood pressure using pulse transit time and pre-ejection
  period,'' in \emph{Convergence Information Technology, 2007. International
  Conference on}.\hskip 1em plus 0.5em minus 0.4em\relax IEEE, 2007, pp.
  941--944.

\bibitem{butlin2016large}
M.~Butlin and A.~Qasem, ``Large artery stiffness assessment using sphygmocor
  technology,'' \emph{Pulse}, vol.~4, no.~4, pp. 180--192, 2016.

\bibitem{ahlstrom2005noninvasive}
C.~Ahlstrom, A.~Johansson, F.~Uhlin, T.~L{\"a}nne, and P.~Ask, ``Noninvasive
  investigation of blood pressure changes using the pulse wave transit time: a
  novel approach in the monitoring of hemodialysis patients,'' \emph{Journal of
  Artificial Organs}, vol.~8, no.~3, pp. 192--197, 2005.

\bibitem{sharwood2005assessment}
G.~Sharwood-Smith, J.~Bruce, and G.~Drummond, ``Assessment of pulse transit
  time to indicate cardiovascular changes during obstetric spinal
  anaesthesia,'' \emph{British Journal of Anaesthesia}, vol.~96, no.~1, pp.
  100--105, 2005.

\bibitem{stiver2017complete}
K.~Stiver, X.~Gao, S.~Shreenivas, K.~Boudoulas, E.~Mazzaferri, N.~Makki, and
  S.~Lilly, ``Complete versus incomplete angiography prior to percutaneous
  coronary intervention in st-elevation myocardial infarction.'' \emph{The
  Journal of invasive cardiology}, 2017.

\bibitem{kubicek1966development}
W.~G. Kubicek, ``Development and evaluation of an impedance cardiac output
  system,'' \emph{Aerospace. Med.}, vol.~37, pp. 1208--1212, 1966.

\bibitem{nyboer1950electrical}
J.~Nyboer, M.~M. Kreider, and L.~Hannapel, ``Electrical impedance
  plethysmography,'' \emph{Circulation}, vol.~2, no.~6, pp. 811--821, 1950.

\bibitem{yamakoshi1980noninvasive}
K.-I. Yamakoshi, H.~Shimazu, T.~Togawa, M.~Fukuoka, and H.~Ito, ``Noninvasive
  measurement of hematocrit by electrical admittance plethysmography
  technique,'' \emph{IEEE Transactions on Biomedical Engineering}, no.~3, pp.
  156--161, 1980.

\bibitem{ghosh2016electrical}
S.~Ghosh, S.~Giri, R.~Kruthika, G.~S. Chabhra, M.~Mahadevappa, and
  J.~Mukhopadhyay, ``Electrical impedance plethysmography based device for
  aortic pulse monitoring,'' in \emph{Systems in Medicine and Biology (ICSMB),
  2016 International Conference on}.\hskip 1em plus 0.5em minus 0.4em\relax
  IEEE, 2016, pp. 124--127.

\bibitem{rezk2011algebraic}
S.~Rezk, C.~Join, and S.~El~Asmi, ``An algebraic derivative-based method for r
  wave detection,'' in \emph{2011 19th European Signal Processing
  Conference}.\hskip 1em plus 0.5em minus 0.4em\relax IEEE, 2011, pp.
  1578--1582.

\bibitem{qasem2008determination}
A.~Qasem and A.~Avolio, ``Determination of aortic pulse wave velocity from
  waveform decomposition of the central aortic pressure pulse,''
  \emph{Hypertension}, vol.~51, no.~2, pp. 188--195, 2008.

\bibitem{galbraith1978coronary}
J.~E. Galbraith, M.~L. Murphy, and N.~de~Soyza, ``Coronary angiogram
  interpretation: interobserver variability,'' \emph{Jama}, vol. 240, no.~19,
  pp. 2053--2056, 1978.

\bibitem{avolio2013arterial}
A.~Avolio, ``Arterial stiffness,'' \emph{Pulse}, vol.~1, no.~1, pp. 14--28,
  2013.

\bibitem{weber2001measurement}
M.~A. Weber, ``The measurement of arterial properties in hypertension,'' 2001.

\bibitem{glaros1988understanding}
A.~G. Glaros and R.~B. Kline, ``Understanding the accuracy of tests with
  cutting scores: The sensitivity, specificity, and predictive value model,''
  \emph{Journal of clinical psychology}, vol.~44, no.~6, pp. 1013--1023, 1988.

\bibitem{paradkar2017coronary}
N.~Paradkar and S.~R. Chowdhury, ``Coronary artery disease detection using
  photoplethysmography,'' in \emph{2017 39th Annual International Conference of
  the IEEE Engineering in Medicine and Biology Society (EMBC)}.\hskip 1em plus
  0.5em minus 0.4em\relax IEEE, 2017, pp. 100--103.

\end{thebibliography}

\end{document}